\documentclass[12pt]{article}
\usepackage{amsmath,amssymb}
\usepackage{bm}
\usepackage{color}
\usepackage{cite}
\usepackage{mathtools}

\usepackage{multirow}

\usepackage{dsfont}

\usepackage{here}

\newcommand{\id}{\mathds{1}}

\begin{document}
		\title{
		\begin{flushright}
			\ \\*[-80pt]
			\begin{minipage}{0.2\linewidth}
				\normalsize
			\end{minipage}
		\end{flushright}
	{$A_4$ Modular Flavour Model of Quark Mass Hierarchies 
		close to the Fixed Point $\tau = \omega$
		\\*[10pt]}}
		\author{
		~S. T.~Petcov $^{1,2}$\footnote{Also at:
			Institute of Nuclear Research and Nuclear Energy,
			Bulgarian Academy of Sciences, 1784 Sofia, Bulgaria.}~  
		and~ M. Tanimoto $^{3}$
		\\*[20pt]
		{
			\begin{minipage}{\linewidth}
				$^1${\it \normalsize
					SISSA/INFN, Via Bonomea 265, 34136 Trieste, Italy} \\*[5pt]
				$^2${\it \normalsize Kavli IPMU (WPI), UTIAS, The University of Tokyo, 
					Kashiwa, Chiba 277-8583, Japan}
				\\*[5pt]
				$^3${\it \normalsize
					Department of Physics, Niigata University, Ikarashi 2, 
					Niigata 950-2181, Japan} \\*[5pt]
			\end{minipage}
		}
		\\*[40pt]}
 \date{
	\centerline{\small \bf Abstract}
	\begin{minipage}{0.9\linewidth}
		\medskip
		\medskip
		\small
		We investigate the possibility to describe the quark mass hierarchies 
		as well as the CKM quark mixing matrix  without 
		fine-tuning in a quark flavour model with modular $A_4$ symmetry. 
		The quark mass hierarchies are considered in the vicinity of the 
		fixed point $\tau = \omega \equiv \exp({i\,2\pi/3})$ (the left cusp of 
		the fundamental domain of the modular group), $\tau$ being the VEV
		of the modulus.  The model involves modular forms of level 3 and weights 
		6, 4 and 2, and contains eight constants, only two of which, $g_u$ 
		and $g_d$, can be a source of CP violation in addition to the VEV of the 
		modulus, $\tau = \omega + \epsilon$, $(\epsilon)^* \neq \epsilon$, 
		$|\epsilon|\ll 1$. We find that in the case of real 
		(CP-conserving) $g_u$ and $g_d$ and common $\tau$ ($\epsilon$) 
		in the down-quark and up-quark sectors, the down-type quark 
mass hierarchies 
		can be reproduced without fine tuning with 
$|\epsilon| \cong 0.03$, all other constants being of the same 
order in magnitude, 
		 and correspond approximately 
		to $1 : |\epsilon| : |\epsilon|^2$. 
		The  up-type quark mass hierarchies
		can be achieved with the same  $|\epsilon| \cong 0.03$ 
		but allowing $g_u\sim {\cal O}(10)$ and correspond to  
		$1 : |\epsilon|/|g_u| : |\epsilon|^2/|g_u|^2$.
		In this setting the reproduction of the value 
		of the CKM element $|V_{\rm cb}|$ is problematic. 
		A much more severe problem is the correct description 
		of the CP violation in the quark sector since it arises 
		as a higher order correction in $\epsilon$ 
with $|\epsilon|\ll 1$.
		We show that this problem is not alleviated in the case of 
		complex $g_u$ and real $g_d$, while in the cases of 
		i) real $g_u$ and complex $g_d$, and  
		ii) complex $g_d$ and  $g_u$, 
the rephasing invariant $J_{\rm CP}$ 
		is larger by a factor of $\sim 1.8$ than the correct value. 
		A correct no-fine-tuned description of the quark mass 
		hierarchies,
		the quark mixing  and CP violation 
		is possible with all constants being of the same 
    order in magnitude and complex $g_u$ and $g_d$, if one allows 
		different values of $\epsilon$ in the down-quark and
 up-quark sectors, 
		or in a modification of the considered model which involves 
		modular forms of level 3 and weights 8, 6 and 4.
	\end{minipage}
}
	
	\begin{titlepage}
		\maketitle
		\thispagestyle{empty}
	\end{titlepage}

\newpage
%
\section{Introduction}
\label{Intro}
%
%

In spite of the remarkable success of the standard model (SM), 
the flavour problem of quarks and leptons is still a challenging issue. 
In order to solve the flavour problem, a remarkable step was made 
in Ref.\cite{Feruglio:2017spp},
where the idea of using modular invariance as a flavour 
symmetry was put forward. 
This new original approach based on modular invariance
opened up a new promising direction in the studies of the flavour physics and correspondingly in flavour model building.

The main feature of the approach proposed in Ref.\cite{Feruglio:2017spp}  
is that the elements of the Yukawa coupling 
and fermion mass matrices in the Lagrangian of the theory are
modular forms of a certain level \(N\) which 
are functions of a single complex scalar field \(\tau\)
-- the modulus -- and have specific transformation properties 
under the action of the modular group.  
In addition, both the couplings and the matter fields
(supermultiplets) are assumed to
transform in representations of an inhomogeneous (homogeneous) 
finite modular group \(\Gamma^{(\prime)}_N\). 
For \(N\leq 5\), the finite modular groups \(\Gamma_N\) 
are isomorphic to the permutation groups 
\( S_3\), \( A_4\), \( S_4\) and \( A_5\)
(see, e.g., \cite{deAdelhartToorop:2011re}), 
while the groups \(\Gamma^\prime_N\) are isomorphic to the double
covers of the indicated permutation groups,
\(S^\prime_3 \equiv S_3\), \(A^\prime_4 \equiv T^\prime\), 
\(S^\prime_4\) and \(A^\prime_5\). These discrete groups 
are widely used in flavour model building.
The theory is assumed to possess the modular symmetry described by the finite modular group \(\Gamma^{(\prime)}_N\), which plays the role of a flavour symmetry.
In the simplest class of such models, 
 the vacuum expectation value (VEV) of  modulus \(\tau\) is the only source of flavour symmetry breaking, such that no flavons are needed. 
 
Another appealing feature of the proposed framework is that 
the VEV of \(\tau\) can also be the only source of breaking of the 
CP symmetry \cite{Novichkov:2019sqv}.
When the  flavour symmetry is broken, 
the elements of the Yukawa coupling and fermion mass matrices
get fixed, and a certain flavour structure arises. 
As a consequence of the modular symmetry, 
in the lepton sector,  for example, 
the charged-lepton and neutrino masses, 
neutrino mixing and the leptonic CPV phases
are simultaneously determined  
in terms of a limited number of coupling constant parameters. 
This together with the fact that they are also functions of 
a single complex VEV -- that of the modulus \(\tau\) --
leads to experimentally testable 
correlations between, e.g., the neutrino mass and 
mixing observables. Models of flavour based on modular invariance 
have then an increased predictive power.

The modular symmetry approach to the flavour problem 
has been widely implemented so far primarily 
in theories with global (rigid) supersymmetry.  
Within the SUSY framework, modular invariance is 
assumed to be a feature of the K\"ahler potential 
and the superpotential of the theory.%
\footnote{Possible non-minimal additions to the K\"ahler potential, 
compatible with the modular symmetry, may jeopardise the predictive power 
of the approach~\cite{Chen:2019ewa}. 
This problem is the subject of ongoing research.
}
Bottom-up modular invariance approaches to the lepton 
flavour problem have been 
exploited first using the groups
\(\Gamma_3 \simeq A_4\)~\cite{Feruglio:2017spp,Criado:2018thu}, 
\(\Gamma_2 \simeq S_3\)~\cite{Kobayashi:2018vbk}, 
\(\Gamma_4 \simeq S_4\)~\cite{Penedo:2018nmg}.
After the first studies, the interest in the approach grew significantly 
and models based on the groups 
\(\Gamma_4 \simeq S_4\)~
\cite{
	Novichkov:2018ovf, Kobayashi:2019mna,Ding:2019gof,  Wang:2019ovr,Wang:2020dbp, Gehrlein:2020jnr},
\(\Gamma_3 \simeq A_4\)~\cite{Kobayashi:2018scp, Novichkov:2018yse, Nomura:2019jxj, Nomura:2019yft, 
	Ding:2019zxk,Nomura:2019lnr,Asaka:2019vev, Zhang:2019ngf, Nomura:2019xsb, Kobayashi:2019gtp, Wang:2019xbo, Abbas:2020vuy, Okada:2020dmb, Ding:2020yen, Behera:2020sfe, Nomura:2020opk,  Behera:2020lpd, Asaka:2020tmo, Okada:2020ukr,Nagao:2020snm, Kobayashi:2021jqu,Okada:2021qdf,Tanimoto:2021ehw,Kobayashi:2021pav,Kobayashi:2022jvy,
	Ahn:2022ufs,Kobayashi:2021ajl,Gunji:2022xig},
\(\Gamma_5 \simeq A_5\)~\cite{Novichkov:2018nkm, Ding:2019xna, Gehrlein:2020jnr},
\(\Gamma_2 \simeq S_3\)~\cite{Okada:2019xqk, Mishra:2020gxg}
and \(\Gamma_7 \simeq PSL(2,\mathbb{Z}_7)\)~\cite{Ding:2020msi}
have been constructed and extensively studied.
Similarly, attempts have been made to construct viable models of 
quark flavour~\cite{Okada:2018yrn} and of quark-lepton unification~\cite{Kobayashi:2018wkl,Okada:2019uoy,Kobayashi:2019rzp,Lu:2019vgm,Abbas:2020qzc,Okada:2020rjb, Du:2020ylx, Zhao:2021jxg,Chen:2021zty,
Ding:2021eva,Ding:2021zbg,Ding:2022bzs,Charalampous:2021gmf,Du:2022lij}.
The formalism of the interplay of modular and generalised CP (gCP) symmetries
has been developed and first applications made in~\cite{Novichkov:2019sqv}. 
It was explored further in~\cite{Okada:2020brs,Yao:2020qyy,Wang:2021mkw,Qu:2021jdy},
as was the possibility of coexistence 
of multiple moduli~\cite{deMedeirosVarzielas:2019cyj, King:2019vhv, deMedeirosVarzielas:2020kji, Ding:2020zxw}, 
considered first phenomenologically 
in~\cite{Novichkov:2018ovf, Novichkov:2018yse}.
Such bottom-up analyses are expected to eventually connect with top-down 
results~\cite{Kobayashi:2018rad, Kobayashi:2018bff, deAnda:2018ecu, Baur:2019kwi, Kariyazono:2019ehj, Baur:2019iai, Nilles:2020nnc, Kobayashi:2020hoc, Abe:2020vmv, Ohki:2020bpo, Kobayashi:2020uaj, Nilles:2020kgo,Kikuchi:2020frp, Nilles:2020tdp, Kikuchi:2020nxn, Baur:2020jwc, Ishiguro:2020nuf, Nilles:2020gvu,  Hoshiya:2020hki, Baur:2020yjl,Liu:2021gwa,Baur:2021bly,
Nilles:2021glx,Kikuchi:2021ogn,Li:2021buv,Kobayashi:2021uam,Hoshiya:2022qvr,Kobayashi:2022sov,Kikuchi:2022svo,Kikuchi:2022geu,Kikuchi:2022pkd,Kikuchi:2021yog}
based on ultraviolet-complete theories.
The problem of modulus stabilisation was also addressed in
\cite{Novichkov:2022wvg,Kobayashi:2019xvz,Kobayashi:2019uyt,Ishiguro:2020tmo}.

While the aforementioned finite quotients~\(\Gamma_N\) of the modular
group have been widely used in the literature to construct modular-invariant
models of flavour from the bottom-up perspective, top-down
constructions typically lead to their double covers~\(\Gamma'_N\) (see,
e.g.,~\cite{Ferrara:1989qb,Baur:2019kwi,Baur:2019iai,Nilles:2020nnc}).
The formalism of such double covers has been developed
first in Refs. \cite{Liu:2019khw}, \cite{Novichkov:2020eep} and 
\cite{Wang:2020lxk,Yao:2020zml} for the cases  
of \(\Gamma'_3 \simeq T'\), \(\Gamma'_4 \simeq S'_4\) and 
\(\Gamma'_5 \simeq A'_5\), respectively, 
where viable lepton flavour models have also been constructed.
Subsequently these groups have been used for flavour model 
building, e.g., in Refs. 
\cite{Liu:2020akv,Okada:2022kee,Ding:2022nzn,Ding:2022aoe,Benes:2022bbg}.

In almost all phenomenologically 
viable flavour models based on modular 
invariance constructed so far
the hierarchy of the charged-lepton and quark 
masses is obtained  by fine-tuning some of the 
constant parameters present in the models.%
\footnote{By fine-tuning we refer to either 
i) unjustified hierarchies between parameters which are introduced in 
the model on an equal footing and/or 
ii) high sensitivity of observables to model parameters.
}
Perhaps, the only notable exceptions are 
Refs.\cite{Criado:2019tzk,King:2020qaj,Kuranaga:2021ujd}, in which modular weights are used 
as Froggatt-Nielsen charges \cite{Froggatt:1978nt}, and additional scalar fields of non-zero modular weights play the role of flavons.
 The recent work in Ref. \cite{Novichkov:2021evw} has proposed the formalism 
that allows to construct models in which 
the fermion (e.g.~charged-lepton and quark) mass hierarchies 
follow solely from the properties of the modular forms, 
thus avoiding the fine-tuning without the need to introduce extra fields.
Indeed, authors have succeeded to reproduce the charged lepton mass hierarchy
without fine-tuning keeping the observed lepton mixing angles.
On the other hand, it is still challenging 
to reproduce quark masses and the Cabibbo, Kobayashi, Maskawa (CKM) 
quark mixing matrix in quark flavour models with modular symmetry.

It was noticed in \cite{Novichkov:2018ovf} 
and further exploited in 
\cite{Novichkov:2018yse,Novichkov:2018nkm,Okada:2020ukr,Okada:2020brs} that  
for the three fixed points of the VEV of \(\tau\) 
in the modular group fundamental domain,
\(\tau_\text{sym} = i\), \(\tau_\text{sym} = 
\omega \equiv \exp(i\,2\pi/ 3)
= -\,1/2 + i\sqrt{3}/2\) (the `left cusp'),
and \(\tau_\text{sym} = i\infty\),
the theories based on the \(\Gamma_N\) 
invariance have respectively 
\(\mathbb{Z}^{S}_2\),
\(\mathbb{Z}^{ST}_3\), and 
\(\mathbb{Z}^{T}_N\) residual symmetries. 
In the case of the double cover groups \(\Gamma^\prime_N\),
the \(\mathbb{Z}^{S}_2\) residual symmetry is replaced by 
the \(\mathbb{Z}^{S}_4\) 
and there is an additional \(\mathbb{Z}_2^R\) symmetry
that is unbroken for any value of \(\tau\)
(see~\cite{Novichkov:2020eep} for further details).

  The fermion mass matrices are strongly constrained
in the points of residual symmetries 
\cite{Novichkov:2018ovf,Novichkov:2018yse,Novichkov:2018nkm,Okada:2020ukr,Okada:2020brs,Novichkov:2021evw,Feruglio:2021dte,Feruglio:2022kea}. 
This suggests that fine-tuning could be avoided in the vicinity 
of these points if the charged-lepton and quark 
mass hierarchies follow from the properties 
of the modular forms present in the corresponding 
fermion mass matrices rather than 
being determined by the values of 
the accompanying constants also present    
in the matrices. 
Relatively small deviations of the modulus VEV 
from the symmetric point might also be needed 
to ensure the breaking of the CP symmetry~\cite{Novichkov:2019sqv}.

\vskip 2mm

In this work, we study the possibility of obtaining the quark mass 
hierarchies as well as the CKM matrix without fine-tuning
along  the lines proposed in Ref. \cite{Novichkov:2021evw} 
in a model with $A_4$ modular quark flavour symmetry.  
Since $A_4$ symmetry is rather simple, it can be used to 
clearly understand the problems facing the 
construction of no-fine-tuned modular invariant 
flavour models of quark mass hierarchies and CKM mixing.
After introducing the necessary tools in Section~\ref{sec:Modular},
we present the $A_4$ modular invariant model in Section \ref{sec:Model}.
In Section \ref{sec:Model}, we describe how one can naturally generate hierarchical mass patterns in the vicinity of symmetric points, and then,
investigate the flavour structure of the quark mass matrices.
In Section \ref{sec:Numerical},  quark masses
and CKM parameters are discussed numerically.
In Section \ref{sec:CPviolation}, the CP problem is discussed.
 We summarize our results  in Section~\ref{sec:Summary}.
 In Appendix \ref{Tensor}, 
 the decomposition of tensor products  are presented.
  In Appendix \ref{Modularforms}, the relevant modular forms with higher weights are listed.
   In Appendix \ref{Nearby}, the modular forms are presented  at close to $\tau=\omega$.
{  In Appendix \ref{fit}, the measure of goodness of
	numerical fitting is presented.}
	
	\section{Modular symmetry of flavours and residual symmetries}
\label{sec:Modular}
%
%
We start by briefly reviewing the modular invariance approach to flavour. 
In this supersymmetric (SUSY) framework, one introduces a chiral superfield, 
the modulus~\(\tau\), transforming non-trivially under the modular group 
\(\Gamma \equiv SL(2, \mathbb{Z})\). The group~\(\Gamma\) is generated by the 
matrices
\begin{equation}
\label{eq:STR_def}
S =
\begin{pmatrix}
0 & 1 \\ -1 & 0
\end{pmatrix}
\,, \quad
T =
\begin{pmatrix}
1 & 1 \\ 0 & 1
\end{pmatrix}
\,, \quad
R =
\begin{pmatrix}
-1 & 0 \\ 0 & -1
\end{pmatrix}\,,
\end{equation}
%
%
obeying \(S^2 = R\), \((ST)^3 = R^2 = \id\), and \(RT = TR\).
The elements \(\gamma\) of the modular group act on \(\tau\) 
via fractional linear transformations,
\begin{equation}
\label{eq:tau_mod_trans}
\gamma =
\begin{pmatrix}
a & b \\ c & d
\end{pmatrix}
\in \Gamma : \quad
\tau \to \gamma \tau = \frac{a\tau + b}{c\tau + d} \,,
\end{equation}
%
while matter superfields transform as ''weighted'' 
multiplets~\cite{Feruglio:2017spp,Ferrara:1989bc,Ferrara:1989qb},
\begin{equation}
\label{eq:psi_mod_trans0}
\psi_i \to (c\tau + d)^{-k} \, \rho_{ij}(\gamma) \, \psi_j \,,
\end{equation}
%
where \(k \in \mathbb{Z}\) is the so-called modular weight%
\footnote{While we restrict ourselves to integer \(k\), it is also possible 
for weights to be fractional 
\cite{Dixon:1989fj,Ibanez:1992hc,Olguin-Trejo:2017zav,Nilles:2020nnc}.
}
and \(\rho(\gamma)\) is a unitary representation of~\(\Gamma\).

In using modular symmetry as a flavour symmetry, an integer level 
\(N \geq 2\) is fixed and one assumes that \(\rho(\gamma) = \id\) 
for elements \(\gamma\) of the principal congruence subgroup
\begin{equation}
\label{eq:congr_subgr}
\Gamma(N) \equiv
\left\{
\begin{pmatrix}
a & b \\ c & d
\end{pmatrix}
\in SL(2, \mathbb{Z}), \,
\begin{pmatrix}
a & b \\ c & d
\end{pmatrix}
\equiv
\begin{pmatrix}
1 & 0 \\ 0 & 1
\end{pmatrix}
(\text{mod } N)
\right\}\,.
\end{equation}
%
Hence, \(\rho\) is effectively a representation of the (homogeneous) 
finite modular group 
\(\Gamma_N' \equiv \Gamma \, \big/ \, \Gamma(N) \simeq SL(2, \mathbb{Z}_N)\). 
For \(N\leq 5\), this group admits the presentation
\begin{equation}
\label{eq:hom_fin_mod_group_pres}
\Gamma'_N = \left\langle S, \, T, \, R \mid S^2 = R, \, (ST)^3 = \id, \, R^2 = \id, \, RT = TR, \, T^N = \id \right\rangle\,.
\end{equation}
%

The  modulus~\(\tau\) acquires a VEV which is restricted to the upper 
half-plane and plays the role of a spurion, parameterising the breaking of 
modular invariance. Additional flavon fields are not required, and we do not 
consider them here. Since~\(\tau\) does not transform under the \(R\) 
generator, a \(\mathbb{Z}_2^R\) symmetry is preserved in such scenarios 
\cite{Novichkov:2020eep}.
If also matter fields transform trivially under \(R\), one may identify 
the matrices \(\gamma\) and \(-\gamma\), thereby restricting oneself to the 
inhomogeneous modular group~\(\overline{\Gamma} \equiv PSL(2, \mathbb{Z}) \equiv SL(2, \mathbb{Z}) \, / \, \mathbb{Z}_2^R\). 
In such a case, \(\rho\) is effectively a representation of a smaller 
(inhomogeneous) finite modular group \(\Gamma_N \equiv \Gamma \, \big/ \left\langle \, \Gamma(N) \cup \mathbb{Z}_2^R \, \right\rangle\). 
For \(N\leq 5\), this group admits the presentation
\begin{equation}
\label{eq:inhom_fin_mod_group_pres}
\Gamma_N = \left\langle S, \, T \mid S^2 = \id, \, (ST)^3 = \id, \, T^N = \id \right\rangle \,.
\end{equation}
%
In general, however, \(R\)-odd fields may be present in the theory and 
\(\Gamma\) and \(\Gamma_N'\) are then the relevant symmetry groups.

Finally, to understand how modular symmetry may constrain the Yukawa couplings 
and mass structures of a model in a predictive way, we turn to the 
Lagrangian -- which for an \(\mathcal{N} = 1\) global supersymmetric theory 
is given by
\begin{equation}
\mathcal{L} = \int \text{d}^2 \theta \, \text{d}^2 \bar{\theta} \, K(\tau,\bar{\tau}, \psi_I, \bar{\psi}_I)
+ \left[ \, \int \text{d}^2 \theta \, W(\tau,\psi_I) + \text{h.c.} \right] \,.
\end{equation}
%
Here \(K\) and \(W\) 
are the K\"ahler potential and the superpotential, respectively. The superpotential \(W\) 
can be expanded in powers of matter superfields \(\psi_I\),
\begin{equation}
\label{eq:W_series}
W(\tau, \psi_I) = \sum \left( \vphantom{\sum} Y_{I_1 \ldots I_n}(\tau) \, \psi_{I_1} \ldots \psi_{I_n} \right)_{\mathbf{1}} \,,
\end{equation}
%
where one has summed over all possible field combinations and 
independent singlets of the finite modular group.
By requiring the invariance of the superpotential under modular 
transformations,
one finds that the field couplings \(Y_{I_1 \ldots I_n}(\tau)\) have to be 
modular forms of level \(N\). These are severely constrained holomorphic 
functions of~\(\tau\), which under modular transformations obey
\begin{equation}
\label{eq:Y_mod_trans}
Y_{I_1 \ldots I_n}(\tau) \,\xrightarrow{\gamma}\, Y_{I_1 \ldots I_n}(\gamma \tau) = (c\tau + d)^{k} \rho_Y(\gamma) \,Y_{I_1 \ldots I_n}(\tau) \,.
\end{equation}
%
Modular forms carry weights \(k = k_{I_1} + \ldots + k_{I_n}\) and furnish 
unitary irreducible representations \(\rho_Y\) of the finite modular group such 
that \(\rho_Y \,\otimes\, \rho_{I_1} \,\otimes \ldots \otimes\, \rho_{I_n} \supset \mathbf{1}\).
Non-trivial modular forms of a given level exist only for  
\(k \in \mathbb{N}\), span finite-dimensional linear spaces
\(\mathcal{M}_{k}(\Gamma(N))\), and can be arranged into multiplets 
of \(\Gamma^{(\prime)}_N\).

The breakdown of modular symmetry is parameterised by the VEV of the modulus 
and there is no value of \(\tau\) which preserves the full symmetry. 
Nevertheless, at certain so-called symmetric points \(\tau = \tau_\text{sym}\) 
the modular group is only partially broken, with the unbroken generators 
giving rise to residual symmetries.
In addition, as we have noticed, the \(R\) generator is unbroken for any 
value of \(\tau\), so that a \(\mathbb{Z}_2^R\) symmetry is always preserved.
There are only three inequivalent symmetric points, namely~
\cite{Novichkov:2018ovf}:
\begin{itemize}
	\item \(\tau_\text{sym} = i \infty\), invariant under \(T\), preserving \(\mathbb{Z}_N^T \times \mathbb{Z}_2^R\);
	\item \(\tau_\text{sym} = i\), invariant under \(S\), preserving \(\mathbb{Z}_4^S \) (recall that \(S^2 = R\)); 
	\item \(\tau_\text{sym} = \omega \equiv \exp (2\pi i / 3)\), `the left cusp', invariant under \(ST\), preserving \(\mathbb{Z}_3^{ST} \times \mathbb{Z}_2^R\).
\end{itemize}

\section{Quark mass hierarchy  in $A_4$ modular invariant  model}
\label{sec:Model}

\subsection{Fermion mass hierarchy without fine-tuning close to $\tau=\omega$}
\label{sec:theory}
%
In theories where modular invariance is broken only by the VEV of modulus, 
the flavour structure of mass matrices in the limit of unbroken supersymmetry 
is determined by the value of \(\tau\) and by the couplings 
in the superpotential.
At a symmetric point \(\tau = \tau_\text{sym}\), flavour textures can be 
severely constrained by the residual symmetry group, which may enforce the 
presence of multiple zero entries in the mass matrices.
As \(\tau\) moves away from its symmetric value, these entries will 
generically become non-zero. The magnitudes of such 
(residual-)symmetry-breaking entries will be controlled by the size of the 
departure \(\epsilon\) from \(\tau_\text{sym}\) and
by the field transformation properties under the residual symmetry group 
(which may depend on the modular weights).
We present below a more detailed discussion of this approach 
to the fermion (charged lepton and quark) mass hierarchies  
following \cite{Novichkov:2021evw}.

Consider a modular-invariant bilinear
\begin{equation}
\label{eq:bilinear}
\psi^c_i \, M(\tau)_{ij}\, \psi_j \,,
\end{equation}
%
where the superfields \(\psi\) and \(\psi^c\) transform under the 
modular group as%
\footnote{Note that in the case of a Dirac bilinear \(\psi\) and \(\psi^c\) 
are independent fields, so in general \(k^c \neq k\) and 
\(\rho^c \neq \rho, \rho^{*}\).}
\begin{equation}
\label{eq:psi_mod_trans}
\begin{split}
\psi \,&\xrightarrow{\gamma}\, (c \tau + d)^{-k} \rho(\gamma) \,\psi \,, \\
\psi^c \,&\xrightarrow{\gamma}\, (c \tau + d)^{-k^c} \rho^c(\gamma)\, \psi^c \,,
\end{split}
\end{equation}
%
so that each \(M(\tau)_{ij}\) is a modular form of level \(N\) and 
weight \(K \equiv k+k^c\).
Modular invariance requires \(M(\tau)\) to transform as
\begin{equation}
\label{eq:mass_matrix_mod_trans}
M(\tau)\, \xrightarrow{\gamma}\, M(\gamma \tau) = (c \tau + d)^K \rho^c(\gamma)^{*} M(\tau) \rho(\gamma)^{\dagger} \,.
\end{equation}
%
Taking \(\tau\) to be close to the symmetric point, and setting \(\gamma\) 
to the residual symmetry generator, one can use this transformation 
rule to constrain the form of the mass matrix \(M(\tau)\). 

Let us  discuss the case, where \(\tau\) is in the vicinity of 
\(\tau_\text{sym} = \omega\). We consider the basis where the 
product \(ST\) is represented by a diagonal matrix.
In this \(ST\)-diagonal basis, 
we  define
\begin{equation}
\label{eq:rhotilde_ST}
\tilde\rho^{(c)}_i \equiv \omega^{k^{(c)}} \rho^{(c)}_i\,,
\end{equation}
%
which are representations under the residual symmetry group.
By setting \(\gamma = ST\) in Eq.\,\eqref{eq:mass_matrix_mod_trans}, 
one finds
\begin{equation}
\label{eq:mass_matrix_ST_trans}
M_{ij}(ST \tau) = [-\omega (\tau + 1)]^K \left( \tilde\rho^c_i \tilde\rho_j \right)^{*} M_{ij}(\tau) \,.
\end{equation}
%
It is now convenient to treat the \(M_{ij}\) as functions of 
\cite{Novichkov:2021evw}
\begin{equation}
u \equiv \frac{\tau - \omega}{\tau - \omega^2} \,, 
\label{eq:u}
\end{equation}
%
so that, in this context, \( |u|\) denotes the deviation of \(\tau\) 
from the symmetric point. 
Note that the entries \(M_{ij}(u)\) depend analytically on \(u\) and that 
\(u \xrightarrow{ST} \omega^2 u\). Thus, in terms of \(u\), 
Eq.\,\eqref{eq:mass_matrix_ST_trans} reads
\begin{equation}
M_{ij}(\omega^2 u) =  \left( \frac{1-\omega^2 u}{1-u} \right)^K
(\tilde\rho^c_i \tilde\rho_j)^{*} M_{ij}(u)
\quad \Rightarrow \quad
\tilde{M}_{ij}(\omega^2 u) = 
(\tilde\rho^c_i \tilde\rho_j)^{*} \tilde{M}_{ij}(u) \,,
\end{equation}
%
where \(\tilde{M}_{ij}(u) \equiv (1-u)^{-K} M_{ij}(u)\). 
Expanding both sides in powers of \(u\), one obtains
\begin{equation}
\label{eq:expansion_ST}
\omega^{2n} \tilde{M}_{ij}^{(n)}(0) = (\tilde\rho^c_i \tilde\rho_j)^{*} \tilde{M}_{ij}^{(n)}(0) \,,
\end{equation}
%
where \(\tilde{M}_{ij}^{(n)}\) denotes the \(n\)-th derivative of \(\tilde{M}_{ij}\) with respect to \(u\).

It follows that for \(\tau \simeq \omega\) the mass matrix entry 
\(M_{ij} \sim \tilde M_{ij}\) is only allowed to be \(\mathcal{O}(1)\) 
when \(\tilde\rho^c_i \tilde\rho_j = 1\). More generally, 
if \(\tilde\rho^c_i \tilde\rho_j = \omega^\ell\) with \(\ell=0,1,2\), 
then the entry \(M_{ij} \sim \tilde M_{ij}\) is expected to be 
\(\mathcal{O}(|u|^\ell)\) in the vicinity of \(\tau = \omega\).
The factors \(\tilde \rho_i^{(c)}\) depend on the weights \(k^{(c)}\), 
see Eq.\,\eqref{eq:rhotilde_ST}. Thus, the leading terms of the  components 
of  the mass matrix is any  of  ${\cal O}(1)$,
${\cal O}(|u|)$ and   ${\cal O}(|u|^2)$ in the vicinity of $\tau=\omega$.
This result 
allows to obtain fermion mass hierarchies without fine-tuning~
\cite{Novichkov:2021evw}.

%
\subsection{Flavour structure of the $A_4$ modular invariant quark model}
\label{sec:Hierarchy}
\subsubsection{The model of quarks}
%
%
We present next a simple model of quark mass matrices 
with modular $A_4$ flavour symmetry, which 
we consider in the vicinity of the fixed point $\tau=\omega$.
We assign  the $A_4$ representation and the weights for the relevant chiral 
superfields of quarks  as
\begin{itemize}
  \item{The three left-handed  doublets  $Q=(Q_1,Q_2,Q_3)$ form a $A_4$ triplet with weight $2$.}
  \item{The  right-handed  singlets $(d^c,s^c,b^c)$ and
  	$(u^c,c^c,t^c)$ are
  	 $A_4$ singlets $(1,\,1'',1')$ with weight $(4,\,2\,,0)$, respectively.}
  \item{The Higgs fields coupled to  up and down sectors  $H_{u},\,H_{d}$ are $A_4$ singlet $1$ with weight 0.}
\end{itemize}
%
\begin{table}[h]
\begin{center}
\renewcommand{\arraystretch}{1.1}
\begin{tabular}{|c|c|c|c|c|} \hline
  & $Q=(Q_1\,,Q_2\,,Q_3)$ & $(d^c,s^c,b^c)$,\
  $(u^c,c^c,t^c)$ &  $H_u$ & $H_d$ \\ \hline
  $SU(2)$ & 2 & 1  & 2 & 2 \\
  $A_4$ & 3 & $(1,\,1'',1')$ & $1$ & $1$ \\
  $k$ & 2 & $(4,\,2,\,0)$ & 0 & 0 \\ \hline
\end{tabular}
\end{center}
\caption{Assignments of $A_4$ representations and weights for relevant chiral super-fields.}
\label{tab:model}
\end{table}
%
These are summarized in Table \ref{tab:model}

\footnote{These assignments of quarks were presented 
in the work of \cite{Okada:2020rjb,Okada:2020ukr}, 
where the quark mass hierarchies are
	obtained by fine-tuning of the constants in the up-quark 
and down-quark mass matrices.}
.
The superpotential terms giving rise to 
quark mass matrices are written by using modular forms with 
weights $2$, $4$ and $6$ as follows: 
\begin{align} 
 &W_d =
\left[\alpha_d ({\bf Y_3^{(6)}}Q)_1 d^c_{1}
 +\alpha'_d ({\bf Y_{3'}^{(6)}}Q)_1 d^c_{1} 
+\beta_d ({\bf Y_3^{(4)}}Q)_{1'}s^c_{1'} 
+ \gamma_d ({\bf Y_3^{(2)}}Q)_{1''}b^c_{1'}\right] H_d\, ,\nonumber \\
&W_u =
\left[\alpha_u ({\bf Y_3^{(6)}}Q)_1 u^c_{1} 
+\alpha'_u ({\bf Y_{3'}^{(6)}}Q)_1 u^c_{1} 
+\beta_u ({\bf Y_3^{(4)}}Q)_{1'}c^c_{1'} 
+ \gamma_u ({\bf Y_3^{(2)}}Q)_{1''}b^t_{1'}\right] H_u\, .
\label{superpotential-I}
\end{align}
%
where the subscripts of $1,1',1''$ denote the $A_4$ representations.
The parameters $\alpha_q$, $\alpha'_q$, $\beta_q$ and  
$\gamma_q$ $(q=d,\,u)$ are, in general, arbitrary complex constants.

We take the modular invariant  kinetic terms simply by 
\begin{equation}
\sum_I\frac{|\partial_\mu\psi^{(I)}|^2}{\langle-i\tau+i\bar{\tau}\rangle^{k_I}} ~,
\label{kinetic}
\end{equation}
%
where $\psi^{(I)}$ denotes a chiral superfield with weight $k_I$,
and $\bar\tau$ is the anti-holomorphic modulus.
After taking VEV  of modulus,  one can set 
$\bar \tau=\tau^*$.
It is important to  address  the transformation needed to get the kinetic 
terms of matter superfields in the canonical form because the
terms  in  Eq.\,(\ref{kinetic}) are not canonical. 
Therefore, we normalize the superfields as:
\begin{eqnarray}
\psi^{(I)}\rightarrow  \sqrt{(2{\rm Im}\tau_q)^{k_I}} \, \psi^{(I)}\,.
\end{eqnarray}
%
The canonical form  is obtained by an overall normalization, which
shifts our parameters  such as
\begin{eqnarray}
&&\alpha_q \rightarrow \hat\alpha_q= \alpha_q\, \sqrt{(2 {\rm Im} \tau_q)^{6} }
=\alpha_q (\sqrt{3}+2 {\rm Im}\,\epsilon)^{3}  ,\, \nonumber\\
&&\alpha'_q \rightarrow \hat\alpha'_q= \alpha'_q\, \sqrt{(2 {\rm Im} \tau_q)^{6} }
=\alpha'_q (\sqrt{3}+2 {\rm Im}\,\epsilon)^{3} \, ,\nonumber\\
&&\beta_q  \rightarrow \hat\beta_q = \beta_q  \, \sqrt{(2 {\rm Im} \tau_q)^{4} }
=\beta_q (\sqrt{3}+2 {\rm Im}\,\epsilon)^{2}\, , \nonumber\\
&&\gamma_q  \rightarrow \hat\gamma_q = \gamma_q \sqrt{(2 {\rm Im} \tau_q)^{2} }
=\gamma_q (\sqrt{3}+2 {\rm Im}\,\epsilon)\, , 
\end{eqnarray}
%
where  $\tau=\omega+\epsilon$ ($|\epsilon|\ll 1$).
We have:
\begin{eqnarray}
\frac{\hat\alpha'_q}{\hat\alpha_q}= \frac{\alpha'_q}{\alpha_q}\,,\quad\frac{\hat\beta_q}{\hat\alpha_q}\simeq \frac{1}{\sqrt{3}}\frac{\beta_q}{\alpha_q}\,,
\qquad \frac{\hat\gamma_q}{\hat\alpha_q}\simeq \frac13\frac{\gamma_q}{\alpha_q}\,.
\end{eqnarray}
%
By using the tensor product decomposition rules
given in Appendix \ref{Tensor}, we obtain the 
following expressions for the mass matrices  $M_d$ and $M_u$ 
of down-type and up-type quarks 
\footnote{We note that the mass matrices are written in RL 
convention.
}:
\begin{align}
&  M_d =v_d
\begin{pmatrix}
\hat\alpha_d  & 0 & 0  \\
0 & \hat\beta_d & 0 \\
0& 0 & \hat\gamma_d \\
\end{pmatrix}
\begin{pmatrix}
\tilde Y_1^{(6)} &  \tilde Y_3^{(6)} &  \tilde Y_2^{(6)} \\
Y_2^{(4)} & Y_1^{(4)} &Y_3^{(4)} \\
Y^{(2)}_3 & Y^{(2)}_2 &Y^{(2)}_1 
\end{pmatrix},\quad
M_u =v_u
\begin{pmatrix}
\hat\alpha_u  & 0  & 0  \\
0 & \hat\beta_u & 0 \\
0& 0 & \hat\gamma_u \\
\end{pmatrix}
\begin{pmatrix}
\tilde Y_1^{(6)} &  \tilde Y_3^{(6)} &  \tilde Y_2^{(6)} \\
Y_2^{(4)} & Y_1^{(4)} &Y_3^{(4)} \\
Y^{(2)}_3 & Y^{(2)}_2 &Y^{(2)}_1 \\
\end{pmatrix},
\label{Massmatrix642}
\end{align}
%
where 
$v_{d(u)}$ denotes VEV  of the neutral component of $H_{d(u)}$, 
\begin{align}
\tilde Y_i^{(6)}=  Y_i^{(6)}+ g_{q} \, Y_i^{'(6)}, \qquad 
g_{q}\equiv\hat\alpha'_q/\hat\alpha_q=\alpha'_q/\alpha_q\,
\qquad (i=1,\,2,\,3,\quad q=d,\,u)\,,
\end{align}
%
and  $Y_i^{(2)}$, $Y_i^{(4)}$ and  $Y_i^{(6)}$ $(i=1,2,3)$,
are the components of the weight 2, 4 and 6 modular forms 
furnishing triplet representations of $A_4$.  
Explicit expressions for the the modular forms of 
interest are presented in Appendix \ref{Modularforms}.

 We note that  
the CKM quark mixing matrix $\rm U_{\rm CKM}$ is given by 
the product of the unitary matrices $\rm U_{uL}$ and $\rm U_{dL}$, which  
diagonalise respectively $M^\dagger_u M_u$ and 
$M^\dagger_dM_d$:  $\rm U_{\rm CKM} = U^\dagger_{uL} U_{dL}$.
It is clear from the expressions of $M_u$ and $M_d$ in 
Eq.\,(\ref{Massmatrix642}) that only the absolute values squared 
of the constants $\alpha_q(\hat\alpha_q )$, $\beta_q(\hat\beta_q )$
and $\gamma_q(\hat\gamma_q )$, $q=d,u$, 
enter into the expressions for $M^\dagger_uM_u$ and $M^\dagger_dM_d$.
Thus, these constants cannot be a source of CP violation and without loss 
of generality can be taken to be real.  
In contrast, the constants $g_q$, $q=d,u$, if complex, may cause 
violation of the CP symmetry and therefore we will consider them, in general,  
as complex parameters.

%
\subsubsection{Quark mass matrices at $\tau=\omega$}
%
%
Consider the quark mass matrices
in Eq.\,\eqref{Massmatrix642}  at the fixed point  $\tau=\omega$.
In the symmetric basis of $S$ and $T$ generators 
given in Appendix \ref{Tensor} (Eq.\,(\ref{ST})), 
in which the mass matrices in Eq.\,(\ref{Massmatrix642}) are obtained,
the modular forms ${\bf Y_3^{(2)}}$, ${\bf Y_3^{(4)}}$ and 
${\bf Y_{3'}^{(6)}}$ take simple forms at $\tau=\omega$ 
as shown in Appendix \ref{Modularforms}. 
Correspondingly, at  $\tau=\omega$ 
the quark mass matrices can written as:
\begin{align}
\begin{aligned}
M_q=
\begin{pmatrix}
-g_{q}\,\hat \alpha_q  & 0 & 0 \\
0 &\hat \beta_q & 0\\
0 & 0 &\hat\gamma_q
\end{pmatrix} 
\begin{pmatrix}
1& -2\omega^2 & -2\omega \\
-\frac{1}{2}\omega &1 &  \omega^2\\
-\frac{1}{2}\omega^2 & \omega& 1
\end{pmatrix}_{RL}\,, \qquad q=d\,, u\,.
\end{aligned}
\label{quark-I}
\end{align}
%
It is easily checked that the 1st, 2nd and 3rd rows
of $M_q$ are proportional to each other.
That is, the mass matrix of Eq.\,\eqref{quark-I} is of rank one.

It proofs convenient to analyse the quark mass matrices $M_q$ 
in the diagonal basis of the $ST$ generator for the $A_4$
triplet, in which the flavour structure of $M_q$ 
becomes explicit.
The $ST$-transformation of the $A_4$ triplet of the left-handed 
quarks $Q$ with weight $k=2$ is
\begin{align}
\begin{pmatrix}
Q_1 \\ Q_2 \\ Q_3\\
\end{pmatrix}
\xrightarrow{ST}
(-\omega-1)^{-2}
\rho(ST) 
\begin{pmatrix}
Q_1 \\ Q_2 \\ Q_3\\
\end{pmatrix}   
= \omega^{2}\times
\frac{1}{3}
\begin{pmatrix}
-1 & 2\omega & 2\omega^2 \\
2 & -\omega & 2\omega^2 \\
2 & 2\omega & -\omega^2 \\
\end{pmatrix}
\begin{pmatrix}
Q_1 \\ Q_2 \\ Q_3\\
\end{pmatrix},
\end{align}
%
where the representations of $S$ and $T$ for the triplet 
are given explicitly in Appendix A.
The $ST$-eigenstate $Q^{ST}$ is obtained 
with the help of a unitary transformation.
We use the unitary matrix $V_{\rm ST}$,
\begin{align}
\begin{aligned}
V_{\rm ST}=\frac{1}{3}
\begin{pmatrix} 
-2 \omega^2  & -2 \omega & 1 \\
\ 2 \omega^2 &- \omega &2\\
- \omega^2  & \ 2 \omega & 2 
\end{pmatrix}\,.
\end{aligned}
\label{STdiagonal}
\end{align}
%
which leads to the diagonal basis of the $ST$ generator of interest:
\begin{align}
\begin{aligned}
V_{\rm ST}\ \omega^{2} {ST} \ V_{\rm ST}^\dagger=
\begin{pmatrix} 
\omega & 0 & 0 \\
0 & 1 & 0 \\
0 &0 &\omega^2
\end{pmatrix}\,, 
\end{aligned} 
\end{align}
%
Then, the $ST$-eigenstate is 
$Q^{ST}= V_{\rm ST}\, Q$.

The right-handed quarks $q_i^c\,(q=d,u)$, which are singlets $(1,1'',1')$
with weights $(4,2,0)$, are eigenstates of $ST$:
\begin{align}
&\begin{pmatrix}
q_1^c \\ q_2^c \\ q_3^c 
\end{pmatrix}
\xrightarrow{ST} 
\begin{pmatrix}
(-\omega-1)^{-4} & 0 & 0 \\
0 & (-\omega-1)^{-2} \,\omega^2 & 0 \\
0 & 0 &\omega \\
\end{pmatrix}
\begin{pmatrix}
q_1^c \\ q_2^c \\ q_3^c \\
\end{pmatrix}
 =
\begin{pmatrix}
\omega & 0 & 0 \\
0 & \omega  & 0 \\
0 & 0 &\omega \\
\end{pmatrix}
\begin{pmatrix}
q_1^c \\ q_2^c \\ q_3^c \\
\end{pmatrix},
\end{align}
%
where we have used $ST$ charges of  $(1,1'',1')$ 
which read $(1,\,\omega^2,\,\omega)$ (see Eq.\,\eqref
{singlet-charge}).
Finally, we have 
\begin{align}
\begin{pmatrix}
u^c \\ c^c \\ t^c 
\end{pmatrix}
\xrightarrow{ST}\  \omega
\begin{pmatrix}
u^c \\ c^c \\ t^c 
\end{pmatrix}\,, \qquad\qquad
\begin{pmatrix}
d^c \\ s^c \\ b^c 
\end{pmatrix}
\xrightarrow{ST}\  \omega
\begin{pmatrix}
d^c \\ s^c \\ b^c 
\end{pmatrix}\,,
\end{align}
%

It can be shown using, in particular, the preceding results 
that  the Dirac mass matrix  
in the $ST$ diagonal basis, 
${\cal M}_{q}$, is related to the mass matrix in the 
initial $S$ and $T$ symmetric basis 
as follows:
\begin{align}
{\cal M}_{q} =  M_{q} V_{ST}^\dagger \,,~\ {\rm q=d,u}\,.
\label{diagonalbase1}
\end{align}
%
Thus, we get:
\begin{align}
{\cal M}^{(0)}_q = M_q V_{\rm ST}^\dagger = 
\begin{aligned}
v_q\,\begin{pmatrix}
0&0 & \frac{27}{8}\,\hat\alpha_q\, g^{}_{q}\, \omega\\
0&0 &  \frac94\, \hat \beta_q\, \omega^2  \\
0 & 0 & \frac32\,\hat\gamma_q \\
\end{pmatrix}
\end{aligned}\,,~\ {\rm q=d,u}\,,
\label{quarkST-I}
\end{align}
%
which is a rank one matrix.
We have two massless up-quarks and two massless 
down-quarks at the fixed point $\tau=\omega$.
The  matrix ${\cal M}_q^\dagger {\cal M}_q$ is transformed  as:
\begin{align}
{\cal M}_q^{(0)\,2}\equiv  V_{\rm ST} M_q^\dagger M_q V_{\rm ST}^\dagger 
= v^2_q\,\frac{9}{4}\,
\begin{aligned}
\begin{pmatrix}
0&0 & 0 \\
0&0 &   0  \\
0 & 0 & \frac{81}{16}\,|g^{\rm }_{q}|^2\,\hat\alpha_q^2+\frac94\,\hat\beta_q^2+\hat\gamma_q^2 \\
\end{pmatrix}
\end{aligned}\,.
\label{quarkST-I2}
\end{align}
%
We see that  only the third generation 
down-types quarks  and up-type ones get non-zero masses.

%
\subsubsection{Quark mass matrices in the vicinity of  $\tau=\omega$}    
%
%

The quark mass matrices in Eq.\,\eqref{quarkST-I}
are corrected due to the small  deviation of $\tau$ 
from the fixed point of $\tau=\omega$.
By the Taylor expansion of the modular forms
in the vicinity of  $\tau=\omega$  as seen  in  Appendix \ref{Nearby}, we estimate
the off-diagonal elements of  ${\cal M}_q^{2}$ 
in Eq.\,(\ref{quarkST-I2}).
In the $ST$ diagonal basis, the correction is parametrised 
by a relatively small variable $\epsilon$, where 
\begin{align}
\tau=\omega+\epsilon\,.
\end{align}
%
The parameter $\epsilon$ 
describing the deviation of $\tau$ from $\omega$ 
is related to the  ``deviation'' parameter $u$ 
introduced in \cite{Novichkov:2021evw} (see Eq.\,(\ref{eq:u})): 
$\epsilon = i\,\sqrt{3}u/(1 - u)\simeq i\,\sqrt{3}u$, 
$|u| \ll 1$.
Up to 2nd order approximation in $\epsilon$,
the quark mass matrix  ${\cal M}_q$ is given by:
\begin{align}
{\cal M}^{(2)}_q =
&v_q
\begin{pmatrix}
\hat\alpha_q\omega Y^3_1  & 0 & 0  \\
0 & \hat\beta_q \omega^2 Y^2_1 & 0 \\
0& 0 & \hat\gamma_q Y_1  \\
\end{pmatrix} \times \nonumber\\
\nonumber\\
&\begin{pmatrix}
(-\,3 + \frac{3}{4} g_q)\epsilon^2_1 & -\,\frac{3}{2}[3\epsilon_1 
+ 2\epsilon^2_1(\frac{5}{2}+\frac{3}{2}k_2)](1+\frac{g_q}{2}) & 
g_q\frac{9}{2}[\frac{3}{4}+\frac{3}{2}\epsilon_1 
+ \epsilon^2_1(\frac{5}{4}+\frac{3}{2}k_2)]
\\
-\,\frac{3}{2}\epsilon^2_1  & \frac{3}{2}[\epsilon_1 + \epsilon^2_1(1+k_2)] & 
\frac{9}{4}+3\epsilon_1 +\epsilon^2_1(\frac{3}{2}+3k_2) \\
\frac{1}{3}\epsilon^2_1 & -\,[\epsilon_1 + \epsilon^2_1(\frac{1}{3}+k_2)] & 
\frac{3}{2}+\epsilon_1 +\epsilon^2_1(\frac{1}{6}+k_2) 
\end{pmatrix}\,,
\label{Massmatrix-I2}
\end{align}
%
where $\epsilon_1$, $\epsilon_2$ , $k_2$ and $k_3$
are given in Appendix \ref{Nearby} :
\begin{align}
\begin{aligned}
\frac{Y_2}{Y_1}\simeq \omega\,(1+\,\epsilon_1 + k_2\epsilon^2_1 ) \, , \qquad 
\frac{Y_3}{Y_1}\simeq -\frac{1}{2}\omega^2 \, 
(1+\, \epsilon_2 + k_3 \epsilon^2_2)\,,
\end{aligned}
\end{align}
%
with
\begin{align}
\epsilon_1 \simeq 2.235\, i\, \epsilon\,,\qquad
\epsilon_2=2\epsilon_1\,,\qquad k_2=0.592\,,
\qquad k_3=0.546\,.
\end{align}
%
Using Eq.\,(\ref{Massmatrix-I2}) we obtain the 
elements of $({\cal M}^{(2)}_q)^\dagger {\cal M}^{(2)}_q \equiv {\cal M}^{2}_q$
in leading order in $\epsilon_1$:
\begin{align}
&{\cal M}_q^{2} [1,1]=
v^2_q\,|\epsilon_1|^4 \left [\frac19|\hat\gamma_q|^2+\frac{9}{16}
(4|\hat\beta_q|^2 + |\hat\alpha_q|^2|g_q - 4|^2)\right ]\,,\nonumber\\
&{\cal M}_q^{2} [2,2]=
v^2_q\, |\epsilon_1|^2 \left [|\hat\gamma_q|^2+\frac{9}{4}
|\hat\beta_q|^2+\frac{81}{16}|\hat\alpha_q|^2|g_q + 2|^2\right ]\,,\nonumber\\
&{\cal M}_q^{2} [3,3]=
 v^2_q\,\left [\frac94|\hat\gamma_q|^2+\frac{81}{16}
|\hat\beta_q|^2+\frac{729}{64}|\hat\alpha_q|^2|g_q|^2\right ]\,,\nonumber\\
&{\cal M}_q^{2} [1,2]= -\,v^2_q\,|\epsilon_1|^2 \epsilon_1^*
\left [\frac13|\hat\gamma_q|^2+ \frac94
|\hat\beta_q|^2-\frac{27}{16}|\hat\alpha_q|^2(2 + g_q)(4 - g_q^*)\right ]\,,\nonumber\\
&{\cal M}_q^{2} [1,3]= v^2_q\, (\epsilon_1^*)^2 
\left [\frac12|\hat\gamma_q|^2- \frac{27}{8}
|\hat\beta_q|^2-\frac{81}{32}|\hat\alpha_q|^2 g_q(4-g_q^*)\right ]\,,\nonumber\\
&{\cal M}_q^{2} [2,3]= -\,v^2_q\, \epsilon_1^*
\left [\frac32|\hat\gamma_q|^2 - \frac98
|\hat\beta_q|^2+\frac{243}{32}|\hat\alpha_q|^2 g_q(2+g_q^*)\right ]\,,\nonumber\\
&{\cal M}_q^{2} [2,1]={\cal M}_q^{2} [1,2]^*\,,\qquad
{\cal M}_q^{2} [3,1]={\cal M}_q^{2} [1,3]^*\,,\qquad
{\cal M}_q^{2} [3,2]={\cal M}_q^{2} [2,3]^*\,,
\label{masscomponent}
\end{align}
%
where the factors $Y_1^3$, $ Y_1^2$ and $Y_1$ in 
$\hat\alpha_q Y_1^3$, $\hat\beta_q Y_1^2$ and $\hat\gamma_q Y_1$
are absorbed in $\hat\alpha_q$, $\hat\beta_q$ and $\hat\gamma_q$,
respectively. We note that in the case of $|\epsilon| \ll 1$ of 
interest the factor $Y_1$  is close to 1 
\footnote{To give a more precise idea of the magnitude of 
$Y_1(\tau)$ we give its value at $\epsilon = 0.0199 + i\,0.02055$, 
which is one of the values of $\epsilon$ relevant for our analysis 
(see further): $Y_1 \simeq 0.95516 -i\,0.00557$.
}. 
The flavour structure of $({\cal M}_q^{(2)})^\dagger {\cal M}_q^{(2)}$ is given
in terms of powers of  $\epsilon$ as:
\begin{align}
{\cal M}_q^{2} \equiv ({\cal M}_q^{(2)})^\dagger {\cal M}_q^{(2)} \sim
v^2_q\,\begin{pmatrix} 
{\cal O}|(\epsilon|^4) & {\cal O}(|\epsilon|^{2}\epsilon^{*}) & {\cal O}(\epsilon^{2*} ) \cr {\cal O}( |\epsilon|^{2} \epsilon^{}) &{\cal O}(|\epsilon|^2)& {\cal O}(\epsilon^*)\cr
{\cal O} (\epsilon^2)  & {\cal O}(\epsilon)& {\cal O}(1)
\end{pmatrix}\,.
\end{align}
%

We can obtain the mass eigenvalues $m_{q1}$, $m_{q2}$ and $m_{q3}$
approximately as follows. The determinant of ${\cal M}_q^{2}$ is given as
\begin{align}
|{\rm det}[{\cal M}_q^{2}]| =m^2_{q1} m^2_{q2} m^2_{q3}
\simeq 729 \,v^6_q\,\hat\alpha_q^2\hat\beta_q^2 \hat\gamma_q^2 
|\epsilon_1|^6\,,
\label{detMq2}
\end{align}
%
which is independent of $g_q$. We also have
\begin{align}
m_{q3}^2\simeq {\cal M}_q^2\,[3,3] =v^2_q\,\frac{9}{64} 
(81\hat\alpha_q^2 |g_q|^2+36\hat\beta_q^2+16\hat\gamma_q^2) \,,
\label{mq32}
\end{align}
%
and
\begin{align}
m_{q2}^2 m_{q3}^2 \simeq v^4_q\,\frac{81}{64} |\epsilon_1|^2
(81\hat\alpha_q^2\hat\beta_q^2 |1+g_q|^2+36\hat\alpha_q^2\hat\gamma_q^2+16\hat\beta_q^2\hat\gamma_q^2) \,.
\label{mq22mq32}
\end{align}
%
It is easy to find that 
in the case of $|g_q|\sim 1$ and 
$\hat\alpha_q\sim \hat\beta_a \sim \hat\gamma_q$ the mass ratios 
satisfy:
\begin{align}
m_{q3} : m_{q2}: m_{q1}\simeq 1:|\epsilon_1|:|\epsilon_1|^2 
\simeq  1:|\epsilon|:|\epsilon|^2\,.
\end{align}
%
On the other hand, if  $|g_q|\gg 1$ and 
$\hat\alpha_q\sim \hat\beta_a \sim \hat\gamma_q$,
we have
 \begin{align}
m_{q3} \simeq \frac{27}{8}\,v_q\,\hat\alpha_q |g_q|  \,,
\quad m_{q2} \simeq 3\,v_q\,\hat\beta_q |\epsilon_1| \,,\quad
 m_{q1}=\frac{8}{3}\,v_q\,\hat\gamma_q|\epsilon_1|^2\frac{1}{ |g_q|}  \,,\quad
\epsilon_1 \simeq 2.235\, i\, \epsilon\,.
\label{dmassratio}
 \end{align}
%
Then, the quark  mass ratios are approximately  given by 
\begin{align}
 m_{q3} : m_{q2}: m_{q1}\simeq \frac{27}{8}|{g_q|}:3|\epsilon_1|:\frac83\frac{1}{{|g_q|}}|\epsilon_1|^2
 = 1 : \frac89 \frac{|\epsilon_1|}{|{g_q|}}: 
  \frac{64}{81}\left (\frac{|\epsilon_1|}{{|g_q|}}\right )^2
  \simeq  1 : \frac{|\epsilon_1|}{|{g_q|}}: 
  \left (\frac{|\epsilon_1|}{{|g_q|}}\right )^2.
  \label{umassratio}
   \end{align}
%
Namely, in the case of  $|g_q|\gg 1$ and 
$\hat\alpha_q\sim \hat\beta_a \sim \hat\gamma_q$
the quark mass hierarchies are given effectively in terms of 
$|\epsilon_1/g_q| \sim |\epsilon/g_q|$.
Indeed, we have succeeded in explaining both down-quark and
up-quark mass hierarchies numerically for $|g_u|\sim 15$ and $|g_d|\sim 1$.
    
Thus, we see the scaling of quark masses with $\epsilon$.
The CKM elements are also scale roughly with  $\epsilon$. However, they depend 
also on the { constants} and phases of both mass matrices because  
both the up and down quark mass matrices contribute to them.
    
%
\section{Quark masses and CKM mixing without fine-tuning}
\label{sec:Numerical}
%
%

 In this Section,   
we discuss the possibility of reproducing the observed 
quark masses and CKM quark mixing parameters 
without fine-tuning in the vicinity of $\tau = \omega$, 
i.e., without strong dependence of the results 
on the constants present in the model.

%
\subsection{Quark mass hierarchies with common $\tau$
in ${\cal M}_d$ and ${\cal M}_u$}
%

We investigate first whether it is possible to describe 
the up-quark and down-quark mass hierarchies 
in terms of powers of the small parameter $\epsilon \equiv \tau-\omega$ 
avoiding fine-tuning of the constants present in the model. 
Correspondingly, we suppose that the constants 
$\alpha_q$, $\alpha'_q$, $\beta_q$ and $\gamma_q$ in 
Eq.\,\eqref{superpotential-I} are real 
and are of the same order, i.e., 
$g_q\equiv\alpha'_q/\alpha_q\simeq\beta_q/\alpha_q\simeq \gamma_q/\alpha_q\sim {\cal O}(1)$, so that their influence on the strong quark 
mass hierarchies of interest is insignificant 
\cite{Novichkov:2021evw}. 
The reality of the constants can be ensured by 
imposing the condition of exact  gCP 
symmetry in the considered model\cite{Novichkov:2019sqv}. The gCP 
symmetry will be broken by the complex value of 
$\epsilon = \tau - \omega \neq 0$  
\footnote{In order for the gCP symmetry to be broken 
the value of $\tau = \omega + \epsilon$ should not lie on the 
border of the fundamental domain of the modular group 
and ${\rm Re}(\tau) \neq 0$ \cite{Novichkov:2021evw}.
}.
It can be broken also by some (or all) constants being complex.

 In the modular invariance approach to the flavour problem 
the modulus $\tau$ obtains a VEV, which breaks the modular flavour 
symmetry, at some high scale. Thus, the quark mass matrices, 
and correspondingly the quark masses, mixing angles and CP violating phase,  
are derived theoretically in the model at this high scale.
The values of these observables at the high scale are 
obtained from the values measured at the electroweak scale 
by the use of the  renormalization group (RG) equations.
In the framework of the minimal SUSY scenario 
the RG running effects depend, in particular,
on the chosen high scale and $\tan\beta$.  
In the analysis which follows we use the GUT scale 
of $2\times 10^{16}$ GeV and $\tan\beta=5$ as reference values.
The numerical values of the quark Yukawa couplings 
at the GUT scale for $\tan\beta=5$ are given by
\cite{Antusch:2013jca, Bjorkeroth:2015ora}:
\begin{align}
\begin{aligned}
&\frac{y_d}{y_b}=9.21\times 10^{-4}\ (1\pm 0.111) \,, 
\qquad
\frac{y_s}{y_b}=1.82\times 10^{-2}\ (1\pm 0.055) \,, \\
\rule[15pt]{0pt}{1pt}
&\frac{y_u}{y_t}=5.39\times 10^{-6}\ (1\pm 0.311)\,, 
\qquad
\frac{y_c}{y_t}=2.80\times 10^{-3}\ (1\pm 0.043)\,.
\end{aligned}\label{Datamass}
\end{align}
%
The quark masses are obtained from the relation 
$m_q=y_q v_H$ with $v_H=174$ GeV.
The choice of relatively small value of $\tan\beta$ allows us to 
avoid relatively large $\tan\beta$-enhanced threshold corrections in 
the RG running of the Yukawa couplings. 
We set these corrections to zero.
 
Assuming that both the ratios of the down-type and up-type 
quark masses, $m_{b}, m_{s}, m_{d}$ and $m_{t}, m_{c}, m_{u}$, 
are determined in the model by the small parameter 
$|\epsilon|$ (or $|\epsilon_1| = 2.235 |\epsilon|$), we have 
\begin{align}
\label{down-hierarchy}
&m_{b} : m_{s}: m_{d}\simeq 1:|\epsilon|:|\epsilon|^2\,,~~
|\epsilon| = 0.02\sim 0.03\,\ (|\epsilon_1| = 0.045\sim 0.067)\,,\\[0.25cm] 
&m_{t} : m_{c}: m_{u} \simeq 1:|\epsilon|:|\epsilon|^2\,,~~
|\epsilon|=0.002\sim 0.003\,\ (|\epsilon_1| = 0.0045\sim 0.0067)\,,
\label{up-hierarchy}
 \end{align}
%
where we have given also the values of  $|\epsilon_1|$
suggested by fitting the down-type and up-type quark mass ratios 
given in  Eq.\,\eqref{Datamass}.   
%
%
%
Thus, the required 
$|\epsilon|$ for the 
description of the down-type and up-type quark 
mass hierarchies differ approximately by one order of 
magnitude. As indicated by Eq.\,\eqref{umassratio}, 
this inconsistency can be ``rescued'' by relaxing the 
requirement on the constant $|g_u|$ in the up-quark sector,  
such as $|g_u|={\cal O}(1)\rightarrow {\cal O}(10)$ leading to
\begin{align}
%
m_{t} : m_{c}: m_{u} \simeq 
1 : \frac{|\epsilon|}{|{g_u|}}: 
\left (\frac{|\epsilon|}{{|g_u|}}\right )^2\,,
\end{align}
%
with $|\epsilon| =0.02\sim 0.03$ 
(corresponding to $|\epsilon_1|= 0.045\sim 0.067$).

 
In the considered case we have eight real parameters in the 
 down-type and up-type quark mass matrices,  
$\alpha_q$, $\beta_q$,  $\gamma_q$, $g_{q}\equiv \alpha'_q/\alpha_q$,
 $q=d,u$, and one complex parameter $\tau = \omega + \epsilon$.
Taking  $\alpha_q\sim\beta_q\sim\gamma_q$, $|g_{d}|= {\cal O}(1) $
and  $|g_{u}|= {\cal O}(10)$, we can reproduce the observed 
quark mass values. A sample set of values of these parameters 
for which the quark mass hierarchies are described correctly  
is given in Tables \ref{tab:input1} and \ref{tab:output1}.
%

 \begin{table}[h]
 	\begin{center}
 		\renewcommand{\arraystretch}{1.1}
 		\begin{tabular}{|c|c|c|c|c|c|c|} \hline
 			\rule[14pt]{0pt}{3pt}  
 			 $\epsilon$ & $\frac{\beta_d}{\alpha_d}$ & $\frac{\gamma_d}{\alpha_d}$ & $g_d$ 
 	 & $\frac{\beta_u}{\alpha_u}$ & $\frac{\gamma_u}{\alpha_u}$ & $g_u$
 	   \\
 	\hline
 	 \rule[14pt]{0pt}{3pt}  
 		 $0.03188+i\, 0.02151$   &$1.69$ & $0.91$  & $-1.94$  & $1.02$
 		 & $0.88$&  $17.83$\\ \hline
 		\end{tabular}
 	\end{center}
 	\caption{ Values of the constant parameters obtained in the fit 
of the quark mass ratios given in Eq.\,\eqref{Datamass}.
}
 	\label{tab:input1}
 \end{table}
  
  \begin{table}[h]
  	\begin{center}
  		\renewcommand{\arraystretch}{1.1}
  		\begin{tabular}{|c|c|c|c|c|} \hline
  			\rule[14pt]{0pt}{3pt}  
  & $\frac{m_s}{m_b}\times 10^2$ & $\frac{m_d}{m_b}\times 10^4$& $\frac{m_c}{m_t}\times 10^3$&$\frac{m_u}{m_t}\times 10^6$
  			\\
  			\hline
  			\rule[14pt]{0pt}{3pt}  
 output &$1.81$ & $ 9.38$
  & $2.77$ & $5.51$\\ \hline
  	\rule[14pt]{0pt}{3pt}
  observed	 &$1.82\pm 0.10$ & $9.21\pm 1.02$ 
  	 & $2.80\pm 0.12$& $ 5.39\pm 1.68$\\ \hline  
  		\end{tabular}
  	\end{center}
  	\caption{
Results on the quark mass ratios compared with 
those at the GUT scale including $1\sigma$ error, given 
in Eq.\,\eqref{Datamass}. 
  }
  	\label{tab:output1}
  \end{table}
%

 A quantitative criterion of fine-tuning, i.e., of
high sensitivity of observables to model parameters,
was proposed by 
R. Barbieri and G. Giudice  in \cite{Barbieri:1987fn}
in a different context, 
but is applicable also in the case of quark mass hierarchies studied by us.
The Barbieri-Giudice measure of fine-tuning~\cite{Barbieri:1987fn} in the 
quark sector, \(\max(\text{BG})\), corresponds to the largest of  quantities 
\( |\partial \ln (\text{mass ratio})/ \partial \ln \tilde\alpha(')_{q}|\), 
\( |\partial \ln (\text{mass ratio})/ \partial \ln \tilde\beta_{q}|\) and 
\( |\partial \ln (\text{mass ratio})/ \partial \ln \tilde\gamma_{q}|\), 
equivalently, to the largest of 
\( |\partial \ln (\text{mass ratio})/ \partial \ln \alpha(')_{q}|\), 
\( |\partial \ln (\text{mass ratio})/ \partial \ln \beta_{q}|\) and 
\( |\partial \ln (\text{mass ratio})/ \partial \ln \gamma_{q}|\). 
An observable \(O\) is typically considered fine-tuned with respect to some 
parameter \(p\) if \(\text{BG} \equiv |\partial \ln O / \partial \ln p| \gtrsim 10\)~\cite{Barbieri:1987fn}.
The criterion is satisfied by the quark mass ratios in the models considered 
in our work. This can be easily checked using the analytic expressions for the 
quark masses in terms of the constant parameters, Eqs. (\ref{detMq2}) 
- (\ref{mq22mq32}) and  (\ref{dmassratio}).
We should add that, as was shown in \cite{Novichkov:2021evw},  
when applied to mixing angles the Barbieri-Giudice criterion leads 
to incorrect results. At present there does not exist a reliable formal 
no-fine-tuning criterion for the 
mixing angles and the CP violating phase. So, we and other authors use 
the simple criterion that the constant parameters present in the 
quark mass matrices be of the same order of magnitude, 
the rational being that these parameters are introduced on equal footing 
and there is no a priori reason why they should have vastly different values. 

%
\subsection{Reproducing the CKM mixing angles}
\label{CKMreal}
%
%

As discussed in the previous sections, the quark mass hierarchies 
are reproduced due to  $\epsilon$, which denotes the deviation from
the  fixed point $\tau=\omega$, and the help of 
$|g_u|\sim {\cal O}(10)$. 
Next, we study the CKM mixing angles by  taking the 
values of $\beta_q/\alpha_q$ ,$\gamma_q/\alpha_q$ and $g_{d}$ to be of order 1.
The present data on the  CKM mixing angles
are given in Particle Data Group (PDG) edition 
of Review of Particle Physics 
\cite{ParticleDataGroup:2022pth} as:
\begin{align}
 \begin{aligned}
 |V_{us}^{\rm }|=0.22500\pm 0.00067 \, , \quad
 |V_{cb}^{\rm }|=0.04182^{\pm 0.00085}_{-0.00074} \,,  \quad
 |V_{ub}^{\rm }|=0.00369\pm 0.00011\, .
 \end{aligned}\label{DataCKM}
 \end{align}
%
By using these values as input and $\tan\beta=5$ we obtain 
the CKM 
mixing angles  at the GUT scale of $2\times 10^{16}$ GeV
\cite{Antusch:2013jca, Bjorkeroth:2015ora}:
  \begin{align}
 |V_{us}^{\rm }|=0.2250\,(1\pm 0.0032) \, , \quad
|V_{cb}^{\rm }|=0.0400\, (1\pm 0.020) \,,  \quad
 |V_{ub}^{\rm }|=0.00353\,(1\pm 0.036)\, .
\label{DataCKM-GUT}
 \end{align}
%
%
%
The tree-level decays of $B\to D^{(*)}K^{(*)}$ are used as the standard candle
of the CP violation.  The CP violating phase of latest world average  
is given in PDG2022 \cite{ParticleDataGroup:2022pth} as:
\begin{equation}
\delta_{CP}={66.2^\circ}^{+ 3.4^\circ}_{-3.6^\circ}\,. 
\label{CKMphase}
\end{equation}
%
Since the phase is almost independent of the evolution of RGE's,
we refer to this value in the numerical discussions.
The rephasing invariant CP violating measure $J_{\rm CP}$ \cite{Jarlskog:1985ht}
is also given in 
\cite{ParticleDataGroup:2022pth}:
\begin{equation}
J_{\rm CP}=3.08^{+0.15}_{-0.13} \times 10^{-5} \,.
\label{JCP}
\end{equation}
%
Taking into account the RG effects on the mixing angles 
for $\tan\beta = 5$, we have at the GUT scale $2\times 10^{16}$ GeV: 
\begin{equation}
J_{\rm CP}= 2.80^{+0.14}_{-0.12}\times 10^{-5}\,.
\label{JCPGUT}
\end{equation}
%
We will discuss the CP violation in our model 
in the next Section.
 

We try to reproduce approximately the observed CKM mixing angles 
with real $g_{q}$, $q=d,u$.
{\ In our scheme, the CKM mixing angles are given 
roughly in terms of powers of 
$\epsilon_1$, as seen in Eq.\,\eqref{masscomponent}. 
In order to reproduce the observed ones precisely, 
the numerical values of the order one ratios of the parameters 
$\beta_q/\alpha_q$ ,$\gamma_q/\alpha_q$ as well as of  
$g_{d}$ ``help'' somewhat (no fine-tuning) since  both up and down 
quark mass matrices contribute to them. We will show those 
numerical values in Tables.
}

We scan parameters with the 
constraint of reproducing the 
observed values of the quark masses,  Cabibbo angle and $|V_{ub}|$ 
including the $3\sigma$ uncertainties. 
A sample set for the  fitting and the results are  
presented in Tables \ref{tab:input2} and  \ref{tab:output2}.
\begin{table}[h]
	\begin{center}
	\renewcommand{\arraystretch}{1.1}
	\begin{tabular}{|c|c|c|c|c|c|c|} \hline
		\rule[14pt]{0pt}{3pt}  
		$\epsilon$ & $\frac{\beta_d}{\alpha_d}$ & $\frac{\gamma_d}{\alpha_d}$ & $g_d$ 
		& $\frac{\beta_u}{\alpha_u}$ & $\frac{\gamma_u}{\alpha_u}$ & $g_u$
		\\
		\hline
		\rule[14pt]{0pt}{3pt}  
		$0.01779+i\, 0.02926$   &$3.26$ & $0.43$  & $-1.40$  & $1.05$
		& $0.80$&  $-16.1$\\ \hline
	\end{tabular}
\end{center}
	\caption{ 
 Values of the constant parameters obtained in the fit of the 
quark mass ratios and of 
CKM mixing angles. See text for details.
}
	\label{tab:input2}
\end{table}

\begin{table}[h]
	\begin{center}
	\renewcommand{\arraystretch}{1.1}
	\begin{tabular}{|c|c|c|c|c|c|c|c|c|} \hline
		\rule[14pt]{0pt}{3pt}  
		& $\frac{m_s}{m_b}\hskip -1 mm\times\hskip -1 mm 10^2$ 
		& $\frac{m_d}{m_b}\hskip -1 mm\times\hskip -1 mm 10^4$& $\frac{m_c}{m_t}\hskip -1 mm\times\hskip -1 mm 10^3$&$\frac{m_u}{m_t}\hskip -1 mm\times\hskip -1 mm 10^6$&
		$|V_{us}|$ &$|V_{cb}|$ &$|V_{ub}|$&$J_{\rm CP}$
		\\
		\hline
		\rule[14pt]{0pt}{3pt}  
		Fit &$1.52$ & $ 8.62$
		& $2.50$ & $5.43 $&
		$0.2230$ & $0.0786$ & $0.00368$ &
		$-2.9\hskip -1 mm\times\hskip -1 mm 10^{-8}$
		\\ \hline
		\rule[14pt]{0pt}{3pt}
		Exp	 &$1.82$ & $9.21$ 
		& $2.80$& $ 5.39$ &
		$0.2250$ & $0.0400$ & $0.00353$ &$2.8\hskip -1 mm\times\hskip -1 mm 10^{-5}$\\
		$1\,\sigma$	&$\pm 0.10$ &$\pm 1.02$ & $\pm 0.12$& $\pm 1.68$ &$ \pm 0.0007$ &
		$ \pm 0.0008$ & $ \pm 0.00013$ &$^{+0.14}_{-0.12}\hskip -1 mm\times \hskip -1 mm 10^{-5}$\\ \hline  
	\end{tabular}
\end{center}
	\caption{Results of the fit of the quark mass ratios, 
CKM mixing angles and $J_{\rm CP}$ factor.
	'Exp' denotes the respective values at the GUT scale, 
including $1\sigma$ errors, quoted in Eqs.\,\eqref{Datamass},
\eqref{DataCKM-GUT} and \eqref{JCPGUT} and obtained from the 
measured ones.
	}
	\label{tab:output2}
\end{table}

As seen in Table \ref{tab:output2},
the values of the CKM elements $|V_{us}|$ and $|V_{ub}|$ 
found in the fit  are consistent with
the observed values. On the other hand, the magnitude of  $V_{cb}$ 
is large, almost twice as large as the observed one.
This result can be understood using 
the results in Eq.\,\eqref{masscomponent}. 
For the values of the parameters in Table \ref{tab:input2} 
both down-type quark sector  and up-type quark one 
contribute to $|V_{cb}|$ additively in ${\cal O}(\epsilon_1)$, 
each contribution being close to the observed one.

The CP violating measure $J_{\rm CP}$ is 
much smaller than the observed one.
As $g_q$ is real, the CP violating phase is generated by 
${\rm Im}\, \epsilon_1$, which corresponds  to $ {\rm Re} \,\epsilon$.
This contribution is strongly suppressed, as discussed 
in Section \ref{sec:CPviolation}. 
%
%

%
\subsection{Reproducing CP violation}
%
%

 As we have seen, the CP violating measure $J_{\rm CP}$ is  
suppressed in the case of real parameters of $g_d$ and  $g_u$ 
due to the extremely small value of the CPV phase $\delta_{\rm CP}$. 
In order to try to reproduce the observed value of $\delta_{\rm CP}$ 
we take  either of the two parameters (or both) to be complex.
In the case of complex  $g_d$ or  $g_u$, the gCP symmetry is broken 
explicitly by the complex constant. 
Having both  $g_d$ and  $g_u$ complex is effectively 
equivalent to not imposing the gCP symmetry requirement at all.
As in the preceding subsection, we scan the parameters with the 
constraint of reproducing the observed values of the quark masses,  
Cabibbo angle and $|V_{ub}|$ including  $3\sigma$ uncertainties.

%
\subsubsection{ Complex $g_d$}
%

First, we take $g_d$ to be complex but  real $g_u$. 
We present a sample set of the results of the fitting 
in Tables \ref{tab:input3} and  \ref{tab:output3}.
We obtain a large value of the CPV phase $\delta_{\rm CP}$
around $80^\circ$. The magnitude of  
$V_{cb}$  is still larger than the observed one by a factor $\sim 2$.
\begin{table}[h]
	\begin{center}
		\renewcommand{\arraystretch}{1.1}
		\begin{tabular}{|c|c|c|c|c|c|c|c|c|} \hline
			\rule[14pt]{0pt}{3pt}  
			$\epsilon$ & $\frac{\beta_d}{\alpha_d}$ & $\frac{\gamma_d}{\alpha_d}$ & $|g_d|$ &${\rm arg}\,[g_d]$ 
			& $\frac{\beta_u}{\alpha_u}$ & $\frac{\gamma_u}{\alpha_u}$ & $|g_u|$&${\rm arg}\,[g_u]$
			\\
			\hline
			\rule[14pt]{0pt}{3pt}  
			$0.00318+i\, 0.02792$   &$2.37$ & $0.41$  & $0.96$ &$161^\circ$  & $1.37$
			& $0.48$&  $15.7$& $0^\circ$\\ \hline
		\end{tabular}
	\end{center}
	\caption{
Values of the constant parameters obtained in the fit of the 
quark mass ratios, CKM mixing angles and of the CPV phase $\delta_{\rm CP}$ 
with complex $\epsilon$ and $g_d$. See text for details.
}
	\label{tab:input3}
\end{table}

\begin{table}[h]
	\small{
		\begin{center}
			\renewcommand{\arraystretch}{1.1}
			\begin{tabular}{|c|c|c|c|c|c|c|c|c|c|} \hline
				\rule[14pt]{0pt}{3pt}  
				& $\frac{m_s}{m_b}\hskip -1 mm\times\hskip -1 mm 10^2$ 
				& $\frac{m_d}{m_b}\hskip -1 mm\times\hskip -1 mm 10^4$& $\frac{m_c}{m_t}\hskip -1 mm\times\hskip -1 mm 10^3$&$\frac{m_u}{m_t}\hskip -1 mm\times\hskip -1 mm 10^6$&
				$|V_{us}|$ &$|V_{cb}|$ &$|V_{ub}|$&$|J_{\rm CP}|$& $\delta_{\rm CP}$
				\\
				\hline
				\rule[14pt]{0pt}{3pt}  
				Fit &$1.53$ & $ 9.41$
				& $2.73$ & $2.33 $&
				$0.2258$ & $0.0775$ & $0.00354$ &
				$5.9\hskip -1 mm\times\hskip -1 mm 10^{-5}$&$81.5^\circ$
				\\ \hline
				\rule[14pt]{0pt}{3pt}
				Exp	 &$1.82$ & $9.21$ 
				& $2.80$& $ 5.39$ &
				$0.2250$ & $0.0400$ & $0.00353$ &$2.8\hskip -1 mm\times\hskip -1 mm 10^{-5}$&$66.2^\circ$\\
				$1\,\sigma$	&$\pm 0.10$ &$\pm 1.02$ & $\pm 0.12$& $\pm 1.68$ &$ \pm 0.0007$ &
				$ \pm 0.0008$ & $ \pm 0.00013$ &$^{+0.14}_{-0.12}\hskip -1 mm\times \hskip -1 mm 10^{-5}$&$^{+ 3.4^\circ}_{-3.6^\circ}$\\ \hline  
			\end{tabular}
		\end{center}
		\caption{Results of the fit of the quark mass ratios, 
CKM mixing angles,  $\delta_{\rm CP}$ and  $J_{\rm CP}$ with complex $\epsilon$ and 
$g_d$. 'Exp' denotes the values of the observables at the GUT scale, 
including $1\sigma$ errors, quoted in Eqs.\,\eqref{Datamass}, 
\eqref{DataCKM-GUT}, \eqref{CKMphase} and \eqref{JCPGUT}  
and obtained from the measured ones.  
		}
		\label{tab:output3}
	}
\end{table}

%
\subsubsection{Complex $g_u$}
%

We consider next  the case of complex $g_u$ and real $g_d$.
A sample set for a  fitting is  in Tables \ref{tab:input4}
and  \ref{tab:output4}. 
The magnitude of  
$V_{cb}$  is also  larger than the observed one by a factor $\sim 2$.
In this case, the magnitude of the CPV phase 
$\delta_{\rm CP}$ is close to  $180^\circ$. Correspondingly, 
the rephasing invariant $|J_{\rm CP}|$ is smaller approximately 
by a factor of 20 than that at the GUT scale, Eq.\,\eqref{JCPGUT}.
\begin{table}[H]
	\begin{center}
		\renewcommand{\arraystretch}{1.1}
		\begin{tabular}{|c|c|c|c|c|c|c|c|c|} \hline
			\rule[14pt]{0pt}{3pt}  
			$\epsilon$ & $\frac{\beta_d}{\alpha_d}$ & $\frac{\gamma_d}{\alpha_d}$ & $|g_d|$ &${\rm arg}\,[g_d]$ 
			& $\frac{\beta_u}{\alpha_u}$ & $\frac{\gamma_u}{\alpha_u}$ & $|g_u|$&${\rm arg}\,[g_u]$
			\\
			\hline
			\rule[14pt]{0pt}{3pt}  
			$0.02344+i\, 0.02510$   &$3.20$ & $0.35$  & $1.41$ &$0^\circ$  & $1.17$
			& $0.32$&  $16.4$& $197^\circ$\\ \hline
		\end{tabular}
	\end{center}
	\caption{ 
Values of the constant parameters obtained in the fit of the 
quark mass ratios, CKM mixing angles and of the CPV phase $\delta_{\rm CP}$ 
with complex $\epsilon$ and $g_u$.
}
	\label{tab:input4}
\end{table}

\begin{table}[h]
	\small{
		\begin{center}
			\renewcommand{\arraystretch}{1.1}
			\begin{tabular}{|c|c|c|c|c|c|c|c|c|c|} \hline
				\rule[14pt]{0pt}{3pt}  
				& $\frac{m_s}{m_b}\hskip -1 mm\times\hskip -1 mm 10^2$ 
				& $\frac{m_d}{m_b}\hskip -1 mm\times\hskip -1 mm 10^4$& $\frac{m_c}{m_t}\hskip -1 mm\times\hskip -1 mm 10^3$&$\frac{m_u}{m_t}\hskip -1 mm\times\hskip -1 mm 10^6$&
				$|V_{us}|$ &$|V_{cb}|$ &$|V_{ub}|$&$|J_{\rm CP}|$& $\delta_{\rm CP}$
				\\
				\hline
				\rule[14pt]{0pt}{3pt}  
				Fit &$1.53$ & $ 7.20$
				& $2.76$ & $2.15 $&
				$0.2248$ & $0.0789$ & $0.00379$ &
				$1.5\hskip -1 mm\times\hskip -1 mm 10^{-6}$&$179^\circ$
				\\ \hline
				\rule[14pt]{0pt}{3pt}
				Exp	 &$1.82$ & $9.21$ 
				& $2.80$& $ 5.39$ &
				$0.2250$ & $0.0400$ & $0.00353$ &$2.8\hskip -1 mm\times\hskip -1 mm 10^{-5}$&$66.2^\circ$\\
				$1\,\sigma$	&$\pm 0.10$ &$\pm 1.02$ & $\pm 0.12$& $\pm 1.68$ &$ \pm 0.0007$ &
				$ \pm 0.0008$ & $ \pm 0.00013$ &$^{+0.14}_{-0.12}\hskip -1 mm\times \hskip -1 mm 10^{-5}$&$^{+ 3.4^\circ}_{-3.6^\circ}$\\ \hline  
			\end{tabular}
		\end{center}
		\caption{Results of the fit of the quark mass ratios, 
CKM mixing angles,  $\delta_{\rm CP}$ and  $J_{\rm CP}$ with complex $\epsilon$ and 
$g_u$. 'Exp' denotes the  values of the observables at the GUT scale, 
including $1\sigma$ error, quoted in Eqs.\,\eqref{Datamass}, 
\eqref{DataCKM-GUT}, \eqref{CKMphase} and \eqref{JCPGUT} 
and obtained from the measured ones. 
		}
		\label{tab:output4}
	}
\end{table}

%
\subsubsection{The case of complex  $g_d$ and $g_u$}
%

Finally, we consider the case of complex $g_d$ and  $g_u$.
A sample set of the results of the fitting is given 
Tables \ref{tab:input5} and  \ref{tab:output5}. 
We obtain a value of the CPV phase $\delta_{\rm CP}$ 
consistent with measured one. On the other hand, the magnitude of  $V_{cb}$  
is still larger than that at the GUT scale given in  Eq.\,\eqref{DataCKM-GUT}.
Therefore, $|J_{\rm CP}|$ is also larger approximately by a factor of 2 
than its value at the GUT scale (see Eq.\,\eqref{JCPGUT}).
It is possible to improve the result for  $|V_{cb}|$, e.g.,  
by  modification of the model, as  is discussed in the next Subsections.
\begin{table}[h]
	\begin{center}
		\renewcommand{\arraystretch}{1.1}
		\begin{tabular}{|c|c|c|c|c|c|c|c|c|} \hline
			\rule[14pt]{0pt}{3pt}  
			$\epsilon$ & $\frac{\beta_d}{\alpha_d}$ & $\frac{\gamma_d}{\alpha_d}$ & $|g_d|$ &${\rm arg}\,[g_d]$ 
			& $\frac{\beta_u}{\alpha_u}$ & $\frac{\gamma_u}{\alpha_u}$ & $|g_u|$&${\rm arg}\,[g_u]$
			\\
			\hline
			\rule[14pt]{0pt}{3pt}  
	$0.00048+i\, 0.02670$   &$2.30$ & $0.39$  & $0.88$ &$161^\circ$  & $1.69$
			& $0.49$&  $16.2$& $205^\circ$\\ \hline
		\end{tabular}
	\end{center}
	\caption{
Values of the constant parameters obtained in the fit of the 
quark mass ratios, CKM mixing angles, the CPV phase $\delta_{\rm CP}$ 
and the $J_{\rm CP}$ factor with complex $\epsilon$, $g_d$ and  $g_u$.
}
	\label{tab:input5}
\end{table}
\begin{table}[H]
	\small{
	\begin{center}
		\renewcommand{\arraystretch}{1.1}
		\begin{tabular}{|c|c|c|c|c|c|c|c|c|c|} \hline
			\rule[14pt]{0pt}{3pt}  
			& $\frac{m_s}{m_b}\hskip -1 mm\times\hskip -1 mm 10^2$ 
			& $\frac{m_d}{m_b}\hskip -1 mm\times\hskip -1 mm 10^4$& $\frac{m_c}{m_t}\hskip -1 mm\times\hskip -1 mm 10^3$&$\frac{m_u}{m_t}\hskip -1 mm\times\hskip -1 mm 10^6$&
			$|V_{us}|$ &$|V_{cb}|$ &$|V_{ub}|$&$|J_{\rm CP}|$& $\delta_{\rm CP}$
			\\
			\hline
			\rule[14pt]{0pt}{3pt}  
			Fit &$1.53$ & $ 8.88$
			& $3.13$ & $2.02 $&
			$0.2229$ & $0.0777$ & $0.00333$ &
			$5.2\hskip -1 mm\times\hskip -1 mm 10^{-5}$&$67.0^\circ$
			\\ \hline
			\rule[14pt]{0pt}{3pt}
			Exp	 &$1.82$ & $9.21$ 
			& $2.80$& $ 5.39$ &
			$0.2250$ & $0.0400$ & $0.00353$ &$2.8\hskip -1 mm\times\hskip -1 mm 10^{-5}$&$66.2^\circ$\\
			$1\,\sigma$	&$\pm 0.10$ &$\pm 1.02$ & $\pm 0.12$& $\pm 1.68$ &$ \pm 0.0007$ &
			$ \pm 0.0008$ & $ \pm 0.00013$ &$^{+0.14}_{-0.12}\hskip -1 mm\times \hskip -1 mm 10^{-5}$&$^{+ 3.4^\circ}_{-3.6^\circ}$\\ \hline  
		\end{tabular}
	\end{center}
	\caption{Results of the fit of the quark mass ratios, 
CKM mixing angles,  $\delta_{\rm CP}$ and  $J_{\rm CP}$ with complex $\epsilon$, 
$g_d$ and $g_u$. 'Exp' denotes the values of the observables at the GUT scale, 
including $1\sigma$ error, quoted in Eqs.\,\eqref{Datamass}, 
\eqref{DataCKM-GUT}, \eqref{CKMphase} and \eqref{JCPGUT} 
and obtained from the measured ones. 
	}
	\label{tab:output5}
}
\end{table}

%
\subsection{CKM mixing angle and CPV phase with two moduli $\tau_d$
and $\tau_u$}
%

In the previous Subsections, we have considered a common modulus $\tau$
in the mass matrices ${\cal M}_d$ and ${\cal M}_u$.
We have seen that the mass hierarchies of both down-type and up-type quarks
can be reproduced with the real constant  $g_u$
having a value $|g_u|$ of ${\cal O}(10)$ and all other constants being 
${\cal O}(1)$. 
We can consider phenomenologically  also the possibility of having 
two different moduli in up- and down- quark sectors, 
$\tau_q$, $q=d,u$ (see Section \ref{sec:CPviolation} for 
further discussion of this possibility).
The d- and u-type quark mass hierarchies in this case are given 
respectively by $1:|\epsilon_u|:|\epsilon_u|^2$ and 
$1:|\epsilon_d|:|\epsilon_d|^2$,
where $|\epsilon_{d,u}|=|\tau_{d,u}-\omega| \ll 1$. 
In this case the  mass hierarchies 
in Eqs.\,\eqref{down-hierarchy} and \eqref{up-hierarchy}
can be easily reproduced with $|g_u|\sim 1$.
\begin{table}[H]
\small{	
		\begin{center}
		\renewcommand{\arraystretch}{1.1}
		\begin{tabular}{|c|c|c|c|c|c|c|c|c|c|} \hline
			\rule[14pt]{0pt}{3pt}  
			$\epsilon_d$	&	$\epsilon_u$ & $\frac{\beta_d}{\alpha_d}$ & $\frac{\gamma_d}{\alpha_d}$ & $|g_d|$ &${\rm arg}\,[g_d]$ 
			& $\frac{\beta_u}{\alpha_u}$ & $\frac{\gamma_u}{\alpha_u}$ & $|g_u|$&${\rm arg}\,[g_u]$
			\\
			\hline
			\rule[14pt]{0pt}{3pt}  
			$0.02331+i\, 0.02269$ 	&	$0.000016+i\, 0.00192$   &$1.43$ & $0.38$  & $1.10$ &$159^\circ$  & $1.58$
			& $1.58$&  $0.895$& $230^\circ$\\ \hline
		\end{tabular}
	\end{center}
	\caption{
Values of the constant parameters obtained in the fit of the 
quark mass ratios, CKM mixing angles and of the CPV phase $\delta_{\rm CP}$ 
in the case of  two moduli $\tau_q = \omega + \epsilon_q$ with complex 
$\epsilon_q$, $q=d,u$,  $\epsilon_d\neq \epsilon_u$.
}
\label{tab:input6}
}
\end{table}

\begin{table}[H]
	\small{
			\begin{center}
			\renewcommand{\arraystretch}{1.1}
			\begin{tabular}{|c|c|c|c|c|c|c|c|c|c|} \hline
				\rule[14pt]{0pt}{3pt}  
				& $\frac{m_s}{m_b}\hskip -1 mm\times\hskip -1 mm 10^2$ 
				& $\frac{m_d}{m_b}\hskip -1 mm\times\hskip -1 mm 10^4$& $\frac{m_c}{m_t}\hskip -1 mm\times\hskip -1 mm 10^3$&$\frac{m_u}{m_t}\hskip -1 mm\times\hskip -1 mm 10^6$&
				$|V_{us}|$ &$|V_{cb}|$ &$|V_{ub}|$&$|J_{\rm CP}|$& $\delta_{\rm CP}$
				\\
				\hline
				\rule[14pt]{0pt}{3pt}  
				Fit &$1.52$ & $ 10.91$
				& $2.71$ & $7.66 $&
				$0.2252$ & $0.0419$ & $0.00351$ &
				$3.2\hskip -1 mm\times\hskip -1 mm 10^{-5}$&$83.8^\circ$
				
\\ 
\hline
\rule[14pt]{0pt}{3pt}
Exp	 &$1.82$ & $9.21$ 
				& $2.80$& $ 5.39$ &
				$0.2250$ & $0.0400$ & $0.00353$ &$2.8\hskip -1 mm\times\hskip -1 mm 10^{-5}$&$66.2^\circ$\\
				$1\,\sigma$	&$\pm 0.10$ &$\pm 1.02$ & $\pm 0.12$& $\pm 1.68$ &$ \pm 0.0007$ &
				$ \pm 0.0008$ & $ \pm 0.00013$ &$^{+0.14}_{-0.12}\hskip -1 mm\times \hskip -1 mm 10^{-5}$&
				$^{+ 3.4^\circ}_{-3.6^\circ}$\\ \hline  
			\end{tabular}
		\end{center}
		\caption{Results of the fits of the quark mass ratios, 
CKM mixing angles, $J_{\rm CP}$ and $\delta_{\rm CP}$ in the vicinity of two different 
moduli in the down-quark and up-quark sectors, $\tau_d$ and $\tau_u$, 
$\tau_d \neq \tau_u$. 'Exp' denotes the  values of the observables 
at the GUT scale, including $1\sigma$ error, quoted in Eqs.\,\eqref{Datamass}, 
\eqref{DataCKM-GUT}, \eqref{CKMphase} and \eqref{JCPGUT} 
and obtained from the measured ones.
		}
		\label{tab:output6}
	}
\end{table}
%

We present a sample set of results of the fitting 
with complex $g_d$ and  $g_u$ 
in Tables \ref{tab:input6} and \ref{tab:output6}.
The magnitude of  $V_{cb}$ 
is almost consistent with the observed one at the GUT scale.
In this case, the up-type quark sector 
contribution to $|V_{cb}|$  is of ${\cal O}(|\epsilon_u|)$, which is 
much smaller than the down-type quark sector 
one that is of ${\cal O}(|\epsilon_d|)$.
{Although we did not search for the  $\chi^2$ minimum,
we show  the magnitude of the measure of goodness of the fitting $N\sigma$,
which is defined  in Appendix \ref{fit}, as a reference value.
In this numerical result, we obtain  $N\sigma=6.8$.
The fit does not look so good.
It is a consequence of the  
extremely high precision of the data 
which we are fitting using simple method 
without making efforts to improve the quality of the fit 
by varying the value of $\tan\beta$ and/or 
the threshold effects in the RG running. 
}

%
\subsection{Improved model with a common modulus $\tau$}	
%
%

Finally, we discuss  an alternative model. In this model we 
introduce  weight 8 modular forms 
in addition to the weights 4 and 6 ones in order 
to get 
a correct description  of the observed three CKM mixing angles 
 and CP violating phase with one modulus $\tau$.
The model is obtained from the considered one by 
replacing the weights  $(4,\,2\,,0)$ of
 the  right-handed  quarks $(d^c,s^c,b^c)$ and
 	$(u^c,c^c,t^c)$  with  weights $(6,\,4\,,2)$, respectively,
in Table \ref{tab:model}.
Then, the quark mass matrices are given as follows:
\begin{align}
&  M_d =v_d
\begin{pmatrix}
\hat\alpha_d  & 0 & 0  \\
0 & \hat\beta_d & 0 \\
0& 0 & \hat\gamma_d \\
\end{pmatrix}
\begin{pmatrix}
\tilde Y_1^{(8)} &  \tilde Y_3^{(8)} &  \tilde Y_2^{(8)} \\
\tilde Y_2^{(6)} &  \tilde Y_1^{(6)} &  \tilde Y_3^{(6)} \\
Y^{(4)}_3 & Y^{(4)}_2 &Y^{(4)}_1 
\end{pmatrix},\quad
M_u =v_u
\begin{pmatrix}
\alpha_u  & 0  & 0  \\
0 & \beta_u & 0 \\
0& 0 & \gamma_u \\
\end{pmatrix}
\begin{pmatrix}
\tilde Y_1^{(8)} &  \tilde Y_3^{(8)} &  \tilde Y_2^{(8)} \\
\tilde Y_2^{(6)} &  \tilde Y_1^{(6)} &  \tilde Y_3^{(6)} \\
Y^{(4)}_3 & Y^{(4)}_2 &Y^{(4)}_1 \\
\end{pmatrix},
\label{Massmatrix-II}
\end{align}
%
where
\begin{align}
\tilde Y_i^{(8)}=  Y_i^{(8)}+ f_{q}  Y_i^{'(8)}\hskip -1 mm,  \ \
\tilde Y_i^{(6)}=  Y_i^{(6)}+ g_{q}  Y_i^{'(6)}\hskip -1 mm,  \ \
f_{q}\equiv\alpha'_q/\alpha_q,\ \, 
g_{q}\equiv\beta'_q/\beta_q\ (i=1,2,3, \, q=d,u)\, .
\end{align}
%
The additional parameters $f_d$ and $f_u$ of the model 
play an important role in reproducing the observed CKM parameters.
Indeed, we have obtained  a good fit of  CKM matrix with 
$|g_d|\simeq |g_u|\simeq |f_d|\simeq 1 $, $|f_u|\simeq 30$ and 
one  $\tau$.
We show the numerical result
with complex $\epsilon$ and   $f_d$, while real $g_d$, $g_u$  and  $f_u$
 in order to reduce free parameters.
The numerical results are presented in Tables \ref{tab:input7} and 
\ref{tab:output7}.
{The measure of goodness of fit is considerably improved as
$N\sigma=1.6$ in this case. 
 We can  get  better  fit with $N\sigma<1$ in the case that
	$g_d$, $g_u$  and  $f_u$ also complex.
	The goodness of the fit might be also  improved by using 
a different value of $\tan\beta$ and/or different set of 
threshold corrections.
}

\begin{table}[h]
	\begin{center}
		\renewcommand{\arraystretch}{1.1} 
		{	\begin{tabular}{|c|c|c|c|c|c|c|c|c|c|} \hline
				\rule[14pt]{0pt}{3pt}  
				$\epsilon$ & $\frac{\beta_d}{\alpha_d}$ & $\frac{\gamma_d}{\alpha_d}$ & $g_d$  & $|f_d|$& ${\rm arg}\,[f_d]$
				& $\frac{\beta_u}{\alpha_u}$ & $\frac{\gamma_u}{\alpha_u}$ & $g_u$
				&$f_u$
				\\
				\hline
				\rule[14pt]{0pt}{3pt}  
				$0.03612+i\, 0.020133$   &$1.78$ & $2.01$  & $-1.43$ &$2.63$ 
				& $78.3^\circ$  & $1.08$
				& $2.11$&  $0.65$&$30.3$ \\ \hline
			\end{tabular}
		}
	\end{center}
	\caption{Values of the constant parameters obtained in the fit of the 
	quark mass ratios, CKM mixing angles, the CPV phase $\delta_{\rm CP}$ 
	and  $J_{\rm CP}$  with complex $\epsilon$ and  $f_d$, while real  $g_d$,  $g_u$  and  $f_u$.
}
	\label{tab:input7}
\end{table}

\begin{table}[H]
	{\small
		\begin{center}
			\renewcommand{\arraystretch}{1.1}
			\begin{tabular}{|c|c|c|c|c|c|c|c|c|c|} \hline
				\rule[14pt]{0pt}{3pt}  
				& $\frac{m_s}{m_b}\hskip -1 mm\times\hskip -1 mm 10^2$ 
				& $\frac{m_d}{m_b}\hskip -1 mm\times\hskip -1 mm 10^4$& $\frac{m_c}{m_t}\hskip -1 mm\times\hskip -1 mm 10^3$&$\frac{m_u}{m_t}\hskip -1 mm\times\hskip -1 mm 10^6$&
				$|V_{us}|$ &$|V_{cb}|$ &$|V_{ub}|$&$|J_{\rm CP}|$& $\delta_{\rm CP}$
				\\
				\hline
				\rule[14pt]{0pt}{3pt}  
				Fit &$1.89$ & $ 9.88$
				& $2.84$ & $3.39 $&
				$0.2250$ & $0.0396$ & $0.00352$ &
				$2.76\hskip -1 mm\times\hskip -1 mm 10^{-5}$&$64.7^\circ$
				\\ \hline
				\rule[14pt]{0pt}{3pt}
				Exp	 &$1.82$ & $9.21$ 
				& $2.80$& $ 5.39$ &
				$0.2250$ & $0.0400$ & $0.00353$ &$2.8\hskip -1 mm\times\hskip -1 mm 10^{-5}$&$66.2^\circ$\\
				$1\,\sigma$	&$\pm 0.10$ &$\pm 1.02$ & $\pm 0.12$& $\pm 1.68$ &$ \pm 0.0007$ &
				$ \pm 0.0008$ & $ \pm 0.00013$ &$^{+0.14}_{-0.12}\hskip -1 mm\times \hskip -1 mm 10^{-5}$&$^{+ 3.4^\circ}_{-3.6^\circ}$\\ \hline  
			\end{tabular}
		\end{center}
	\caption{Results of the fit of the quark mass ratios, 
		CKM mixing angles,  $\delta_{\rm CP}$ and  $J_{\rm CP}$  with complex $\epsilon$ and  $f_d$, while real $g_d$, $g_u$  and  $f_u$. 'Exp' denotes the values of the observables at the GUT scale, 
		including $1\sigma$ error, quoted in Eqs.\,\eqref{Datamass}, 
		\eqref{DataCKM-GUT}, \eqref{CKMphase} and \eqref{JCPGUT} 
		and obtained from the measured ones. 
	}
		\label{tab:output7}
	}
\end{table}



%
\section{The CP Violation Problem}
\label{sec:CPviolation}
%
The mass matrices ${\cal M}_q$, $q=d,u$, we have obtained, 
Eq.\,\eqref{Massmatrix-I2}, are 
written in the $RL$ basis of the right-handed and left-handed quark fields.
They are diagonalised by the bi-unitary transformations:
$\rm {\cal M}_q = U_{qR} {\cal M}_q^{diag} U^\dagger_{qL}$, 
where $\rm {\cal M}^{diag}_{d,u}$ are diagonal matrices with the masses of 
$d,s,b$ and $u,c,t$   quarks, and $\rm U_{qR}$ and $\rm U_{qL}$ 
are unitary matrices. The CKM quark mixing 
matrix $\rm U_{CKM}$ is given by  $\rm U_{CKM}= U^\dagger_{uL}U_{dL}$.
The matrices  $\rm U_{dL}$ and $\rm U_{uL}$ diagonalise 
${\cal M}_d^\dagger{\cal M}_d$ and ${\cal M}_u^\dagger{\cal M}_u$, 
respectively: ${\cal M}_q^\dagger {\cal M}_q = 
{\rm U_{qL}}({\cal M}_q^{diag})^2 {\rm U^\dagger_{qL}}$.  
Thus, the possibility to have CP-violating  $\rm U_{CKM}$, 
as required by the data, is determined by whether 
${\cal M}_q^\dagger {\cal M}_q$ violate the CP symmetry or not, 
i.e., by whether ${\cal M}_q^\dagger {\cal M}_q$ contain complex elements 
that break the CP symmetry. 

We note first that ${\cal M}_q^\dagger{\cal M}_q$ depend on 
$|\hat\alpha_q\omega Y^3_1|^2$, $|\hat\beta_q \omega^2 Y^2_1|^2$ 
and $|\hat\gamma_q Y_1|^2$, and therefore the constants 
$\hat\alpha_q$, $\hat\beta_q$, $\hat\gamma_q$ and $Y_1$ 
cannot be a source of CP violation.

Suppose next that the constants $g_q$, $q=d,u$, present in the expression 
Eq.\,(\ref{Massmatrix-I2}) of ${\cal M}_q$ are real and that 
$\epsilon$ is complex such that 
$\tau = \omega + \epsilon$ is CP-violating 
\footnote{We recall that $\epsilon_1 \simeq i\,2.235 \epsilon$
	in Eq.\,\eqref{ep1ep2} of Appendix \ref{Nearby}.}, 
i.e., that $\tau$ is the only source of breaking of 
CP symmetry \cite{Novichkov:2019sqv}\,\footnote{The reality of the constants present in the model 
	can be ensured by imposing the gCP symmetry 
	\cite{Novichkov:2019sqv}.
}.
The deviation $\epsilon$ of $\tau$ from the left cusp 
plays the role of a small parameter in terms of which the 
quark mass hierarchies are expressed. 
We have found that in this case the correct description 
of the down-quark and up-quark mass hierarchies, which are very different,
can be achieved with the help of the real constant $g_u$ having a 
relatively large  absolute value,  $|g_{u}|= {\cal O}(10)$.
which provides the necessary ``enhancement'' of the up-quark 
mass hierarchies, with all the other constants being of the same order 
in magnitude and much smaller than $g_u$.
The mass hierarchies thus obtained have the following 
forms: 
\begin{align}
\label{dQMH}
& m_b:m_s:m_d \sim m_b\,(1:|\epsilon|:|\epsilon|^2)\,, \\ 
\label{uQMH}
& m_t:m_c:m_u \sim m_t\,(1:|\epsilon/g_u|:|\epsilon/g_u|^2)\,.
\end{align}
%
For the discussion of the problem of CP violation, however,
the presence of the real constant $g_u$ in Eq.\,(\ref{uQMH}) 
and in the up-quark mass matrix  ${\cal M}_u$
is not relevant and for simplicity of the presentation 
we will refer in what follows in this Section to both 
hierarchies in Eqs.\,(\ref{dQMH}) and (\ref{uQMH}) 
as being of the type $1:|\epsilon|:|\epsilon|^2$.
Then from the point of view of CP violation 
the quark mass matrices  ${\cal M}_{d,u}$ in 
Eq.\,(\ref{Massmatrix-I2}) 
have the following generic structure:
\begin{align}
&{\cal M}^{gen}_q =v_q
\begin{pmatrix}
i^2\,\epsilon^2 & i\,\epsilon & 1 \\
i^2\,\epsilon^2 & i\,\epsilon & 1 \\
i^2\,\epsilon^2 & i\,\epsilon & 1 
\end{pmatrix}\,,~~q=d,u\,,
\label{MassMIgeneric}
\end{align}
%
where we have 
used $\epsilon_1 =i\,2.235\epsilon$ and  
kept only the leading order terms in $\epsilon$ 
in the first, second and third columns of ${\cal M}_q$.
The different real coefficients in the elements of 
 ${\cal M}_q$ shown in the second matrix in 
Eq.\,(\ref{Massmatrix-I2}), including the factors 
$(2.235)^2$ and 2.235 in the first and second column, 
as well as the common factors of the 1st, 2nd and 3rd rows of  
${\cal M}_q$, namely, $\hat\alpha_q\omega Y^3_1$,
$\hat\beta_q \omega^2 Y^2_1$ and $\hat\gamma_q Y_1$,  
which we have  not included in the expression 
Eq.\,(\ref{MassMIgeneric}) of ${\cal M}_q$, are not relevant for 
the present general discussion of CP violation. 
It is not difficult to show that $({\cal M}^{gen}_q)^\dagger
{\cal M}^{gen}_q$ 
of interest can be cast in the form:
\begin{align}
&({\cal M}^{gen}_q)^\dagger{\cal M}^{gen}_q  =v^2_q
\begin{pmatrix}
-\,i\,e^{-i\,\kappa_q}  & 0 & 0  \\
0 & 1 & 0 \\
0& 0 & i\,e^{i\,\kappa_q} 
\end{pmatrix} \hskip -2 mm
\begin{pmatrix}
|\epsilon_{q}|^4 & |\epsilon_{q}|^3 & |\epsilon_{q}|^2 \\
|\epsilon_{q}|^3 & |\epsilon_{q}|^2 & |\epsilon_{q}| \\
|\epsilon_{q}|^2 & |\epsilon_{q}| & 1 
\end{pmatrix}\hskip -2 mm
\begin{pmatrix}
i\,e^{i\,\kappa_q}  & 0 & 0  \\
0 & 1 & 0 \\
0& 0 & -\,i\,e^{-i\,\kappa_q} 
\end{pmatrix}, ~q=d,u,
\label{MgenDagMgen}
\end{align}
%
where  we have used 
$\epsilon_{q} = |\epsilon_{q}|e^{i\,\kappa_q}$ 
taking into account the possibility of two 
different deviations from $\tau = \omega$ 
in the down-quark and up-quark sectors
\footnote{
	We have omitted also the overall factor 3 
	in the right-hand side of Eq.\,(\ref{MgenDagMgen}),
	which is irrelevant for the current discussion.
}.
The matrix in Eq.\,(\ref{MgenDagMgen}) 
is diagonalised by $\rm U^{gen}_{qL} = P(\kappa_q)O_q$,
where $\rm P(\kappa_q) = 
{\rm diag}(e^{-i\,(\kappa_q+\pi/2)},1, e^{i\,(\kappa_q+\pi/2)})$ 
and $\rm O_q$ is a real orthogonal matrix.
The CKM matrix in this schematic analysis of ``leading order'' 
CP violation is given by: 
\begin{equation}
\rm U^{\rm gen}_{\rm CKM} = O^T_u\,P^*(\kappa_u)P(\kappa_d)O_d\,.
\label{UgenCKM}
\end{equation}
%

In the ``minimal'' case of one and the same deviation of 
$\tau$ from $\omega$ we have $\epsilon_{1d} = \epsilon_{1u}$   
and, consequently, $\kappa_d = \kappa_u$.
It follows from Eq.\,(\ref{UgenCKM}) that in this case 
$\rm U^{gen}_{\rm CKM}$ is real and thus CP conserving.
This implies that to leading order in $\epsilon$ 
in the elements of the quark mass matrices 
${\cal M}_q$, there will be no CP violation in the quark 
sector in the considered model: the CP violating phase 
in $\rm U_{\rm CKM}$, $\delta^{th}_{\rm CP} = 0,\pi$, 
while it follows from the data that  $\delta_{\rm CP} \simeq 66.2^\circ$.
The CP violation arises as a higher order 
due to the corrections to the leading terms in the elements 
of ${\cal M}_q$.
%
 Since it is possible to describe correctly the quark mass hierarchies only if 
we have $|\epsilon| \ll 1$, 
the CPV phase $\delta^{th}_{\rm CP}$, 
which is generated by the higher order corrections in $\epsilon$ 
in the elements of ${\cal M}_q$,   
is generically much smaller than the 
measured value of $\delta_{\rm CP}$, i.e., 
$\delta^{th}_{\rm CP} \ll 66.2^\circ$, which is incompatible 
with the data. This conclusion is confirmed by our numerical 
analysis in Section \ref{CKMreal}. 

Thus, we arrive at the conclusion that in the 
considered quark flavour model with $A_4$ modular 
symmetry, supplemented by the gCP symmetry,
and one modulus $\tau$ having a VEV 
in the vicinity of the left cusp, $\tau = \omega + \epsilon$,
the description of the quark mass hierarchies in terms of 
$\epsilon$, which has the generic structure  
$1:|\epsilon|:|\epsilon|^2$ and thus implies $|\epsilon| \ll 1$,
is incompatible with the description of CP violation in the quark sector.
On the basis of the general results presented in \cite{Novichkov:2021evw}
we suppose that the problem of incompatibility 
between the ``no-fine-tuned'' description of the quark mass hierarchies 
in the vicinity of the left cusp $\tau = \omega + \epsilon$ 
with $|\epsilon|\ll 1$, and the description of 
CP violation in the quark sector, will be 
present in any quark flavour model 
based on the finite modular groups 
$S_3$, $A^{(\prime)}$, $S^{(\prime)}_4$ and $A^{(\prime)}_5$
and gCP symmetry.

In the modular $A_4$ model 
studied by us we have considered 
phenomenologically also the possibility of having 
two different moduli in down- and up-type quark sectors, 
$\tau_q$, $q=d,u$, acquiring VEVs in the vicinity of the 
left cusp, $\tau_q = \omega + \epsilon_q$, $q=d,u$,  
with $\epsilon_d \neq \epsilon_u$ 
\footnote{Models based on simplectic modular groups contain naturally 
	more than one modulus \cite{Ding:2021iqp}. Constructing a self-consistent 
	quark flavour model with two moduli is beyond the scope of the present 
	study.}.
The d- and u-quark mass hierarchies in this case are given 
respectively by $1:|\epsilon_d|:|\epsilon_d|^2$ and 
$1:|\epsilon_u|:|\epsilon_u|^2$, $|\epsilon_{d,u}| \ll 1$. 
Since  $\epsilon_d \neq \epsilon_u$, we have also 
$\kappa_d \neq \kappa_u$ and thus 
$\rm P^*(\kappa_u)P(\kappa_d) = 
diag(e^{i\,(\kappa_u - \kappa_d)},1, e^{-i\,(\kappa_u - \kappa_d)}) \neq \id$ 
in Eq.\,(\ref{UgenCKM}), 
where $\id$ is the unit matrix. The factor 
$\rm P^*(\kappa_u)P(\kappa_d)$
may, in principle, be a 
source of the requisite CP violation provided $\kappa_u - \kappa_d$ 
is sufficiently large. However, performing a numerical analysis 
we  find that the CPV phase $\delta_{\rm CP}$
thus generated is too small to be compatible with the measured value.
Thus, the problem of correct description of the CP violation in the quark 
sector in the model considered by us persist also in this case as long as 
the gCP symmetry constraint is imposed. 

%
\section{Summary}
\label{sec:Summary}
%

 We have investigated the possibility to describe the quark mass hierarchies 
as well as the CKM quark mixing matrix  without 
fine-tuning in a quark flavour model with modular $A_4$ symmetry. 
The quark mass hierarchies are considered in the vicinity of the 
fixed point $\tau = \omega \equiv \exp({i\,2\pi/3})$ (the left cusp of 
the fundamental domain of the modular group), $\tau$ being the VEV 
of the modulus. In the considered $A_4$ model 
the three left-handed (LH) quark doublets $Q = (Q_1,Q_2,Q_3)$ are assumed to 
furnish a triplet irreducible representation of $A_4$ 
and to carry weight 2, while the three right-handed (RH) up-quark and 
down-quark singlets are supposed to be the $A_4$ singlets 
$({\bf 1},{\bf 1''},{\bf 1'})$ carrying weights (4,2,0), respectively.
The model involves modular forms of level 3 and weights 
6, 4 and 2, and contains eight constants, only two of which, $g_d$ 
and $g_u$, can be a source of CP violation in addition to the VEV of the 
modulus, $\tau = \omega + \epsilon$, $(\epsilon)^* \neq \epsilon$, 
$|\epsilon|\ll 1$.

 We find that in the case of real 
(CP-conserving) $g_d$ and $g_u$ and common $\tau$ ($\epsilon$) 
for the down-type quark and up-type quark sectors, the down-type quark mass hierarchies 
can be reproduced without fine tuning with $|\epsilon| \cong 0.03$, 
all other constants being of the same order in magnitude, and correspond 
approximately 
to $1 : |\epsilon| : |\epsilon|^2$. The description of the up-type 
quark mass hierarchies requires a ten times smaller value of 
$|\epsilon|$. It can be achieved with the same  $|\epsilon| \cong 0.03$ 
allowing the constant $g_u$ to be larger in magnitude 
than the other constants of the model, $|g_u|\sim {\cal O}(10)$, 
and corresponds to  $1 : |\epsilon|/|g_u| : |\epsilon|^2/|g_u|^2$.
In this setting the description of the CKM element $|V_{\rm cb}|$
is problematic. We have shown that a much more severe problem 
is the correct description of the CP violation 
in the quark sector since it arises as a higher order correction 
in $\epsilon$, which has to be sufficiently small in order to 
reproduce the quark mass hierarchies. 
This problem may be generic to modular invariant quark flavour models 
with one modulus, in which the gCP symmetry is imposed and 
the quark mass hierarchies are obtained   
in the vicinity of fixed point $\tau = \omega$. 
In the considered  model  the CP violation problem is not alleviated 
in the case of complex $g_u$ and real $g_d$, while in the cases of 
i) real $g_u$ and complex $g_d$, and  
ii) complex $g_d$ and  $g_u$, the rephasing invariant $J_{\rm CP}$ 
has a value which is larger by a factor of $\sim 1.8$ than the correct 
value due to a larger than observed value of $|V_{\rm cb}|$.
We show also that an essentially correct description of the quark mass 
hierarchies and the CKM mixing matrix, including the CP violation 
in the quark sector, is possible in the vicinity of the left cusp
with all constants being of $\sim {\cal O}(1)$ in 
magnitude and complex $g_d$ and $g_u$, if there are two different moduli 
$\tau_d = \omega + \epsilon_d$ and $\tau_u = \omega + \epsilon_u$ 
in the down-quark and up-quark sectors, with down-type and up-quark mass 
hierarchies given by $1 : |\epsilon_d| : |\epsilon_d|^2$ and 
$1 : |\epsilon_u| : |\epsilon_u|^2$. A correct description 
is also possible in a modification of the considered model 
which involves level 3 modular forms of weights 8, 6 and 4.
Thus, ten observables of quark sector can be reproduced quite well.
	On the other hand, there is no room to predict somethings 
	in the quark masses and mixings. 
{ However, the model has predictive power for the flavor phenomena 
in SMEFT, such as in  B meson decays
as well as the  flavor violation of the charged lepton decays
\cite{Kobayashi:2022jvy}.
}

 The results of our study show that describing correctly 
without sever fine-tuning the quark mass hierarchies, 
the quark mixing and the CP violation in the quark sector 
is remarkably challenging within the modular invariance approach 
to the quark flavour problem.

\section*{Acknowledgments}
The work of S. T. P. was supported in part by the European
Union's Horizon 2020 research and innovation programme under the 
Marie Sk\l{}odowska-Curie grant agreement No.~860881-HIDDeN, by the Italian 
INFN program on Theoretical Astroparticle Physics and by the World Premier 
International Research Center Initiative (WPI Initiative, MEXT), Japan.
The authors would like to thank Kavli IPMU, University of Tokyo, 
where part of this study was performed, for the kind hospitality. 

%
\appendix
\section*{Appendix}
%
%
\section{Tensor product of  $\rm A_4$ group}
\label{Tensor}
%

We take the generators of $A_4$ group for the triplet as follows:
\begin{align}
\begin{aligned}
S=\frac{1}{3}
\begin{pmatrix}
-1 & 2 & 2 \\
2 &-1 & 2 \\
2 & 2 &-1
\end{pmatrix},
\end{aligned}
\qquad 
\begin{aligned}
T=
\begin{pmatrix}
1 & 0& 0 \\
0 &\omega& 0 \\
0 & 0 & \omega^2
\end{pmatrix}, 
\end{aligned}
\label{ST}
\end{align}
%
where $\omega=e^{i\frac{2}{3}\pi}$.
In this basis, the multiplication rules are:
\begin{align}
& \begin{pmatrix}
a_1\\
a_2\\
a_3
\end{pmatrix}_{\bf 3}
\otimes 
\begin{pmatrix}
b_1\\
b_2\\
b_3
\end{pmatrix}_{\bf 3} 
\nonumber \\
&=\left (a_1b_1+a_2b_3+a_3b_2\right )_{\bf 1} 
 \oplus \left (a_3b_3+a_1b_2+a_2b_1\right )_{{\bf 1}'}
 \oplus \left (a_2b_2+a_1b_3+a_3b_1\right )_{{\bf 1}''} 
\nonumber \\
&\oplus \frac13
\begin{pmatrix}
2a_1b_1-a_2b_3-a_3b_2 \\
2a_3b_3-a_1b_2-a_2b_1 \\
2a_2b_2-a_1b_3-a_3b_1
\end{pmatrix}_{{\bf 3}}
\oplus \frac12
\begin{pmatrix}
a_2b_3-a_3b_2 \\
a_1b_2-a_2b_1 \\
a_3b_1-a_1b_3
\end{pmatrix}_{{\bf 3}\  } \ ,\\
& {\bf 1} \otimes {\bf 1} = {\bf 1}\,, \qquad 
{\bf 1'} \otimes {\bf 1'} = {\bf 1''}\,, \qquad
{\bf 1''} \otimes {\bf 1''} = {\bf 1'}\,, \qquad
{\bf 1'} \otimes {\bf 1''} = {\bf 1}\,,
\nonumber \\
& {\bf 1'} \otimes
\begin{pmatrix}
a_1\\
a_2\\
a_3
\end{pmatrix}_{\bf 3}
= 
\begin{pmatrix}
a_3\\
a_1\\
a_2
\end{pmatrix}_{\bf 3}\,, \qquad 
{\bf 1''}\otimes
\begin{pmatrix}
a_1\\
a_2\\
a_3
\end{pmatrix}_{\bf 3}
= 
\begin{pmatrix}
a_2\\
a_3\\
a_1
\end{pmatrix}_{\bf 3}\,, 
\end{align}
%
where
\begin{align}
S({\bf 1')}=1\,,\qquad S({\bf 1''})=1, \qquad T({\bf 1')}=\omega\,,\qquad T({\bf 1''})=\omega^2. 
\label{singlet-charge}
\end{align}
%
Further details can be found in the reviews~
\cite{Ishimori:2010au,Ishimori:2012zz,Kobayashi:2022moq}.

%
\section{Modular forms of $A_4$}
\label{Modularforms}
%
%
The  modular forms of weight $2$ transforming
as a triplet of $A_4$ can be written in terms of 
$\eta(\tau)$ and its derivative \cite{Feruglio:2017spp}:
\begin{eqnarray} 
\label{eq:Y-A4}
Y_1 &=& \frac{i}{2\pi}\left( \frac{\eta'(\tau/3)}{\eta(\tau/3)}  +\frac{\eta'((\tau +1)/3)}{\eta((\tau+1)/3)}  
+\frac{\eta'((\tau +2)/3)}{\eta((\tau+2)/3)} - \frac{27\eta'(3\tau)}{\eta(3\tau)}  \right), \nonumber \\
Y_2 &=& \frac{-i}{\pi}\left( \frac{\eta'(\tau/3)}{\eta(\tau/3)}  +\omega^2\frac{\eta'((\tau +1)/3)}{\eta((\tau+1)/3)}  
+\omega \frac{\eta'((\tau +2)/3)}{\eta((\tau+2)/3)}  \right) , \label{Yi} \\ 
Y_3 &=& \frac{-i}{\pi}\left( \frac{\eta'(\tau/3)}{\eta(\tau/3)}  +\omega\frac{\eta'((\tau +1)/3)}{\eta((\tau+1)/3)}  
+\omega^2 \frac{\eta'((\tau +2)/3)}{\eta((\tau+2)/3)}  \right)\,,
\nonumber
\end{eqnarray}
%
which satisfy also the constraint \cite{Feruglio:2017spp}:
\begin{align}
Y_2^2+2Y_1Y_3=0~.
\label{condition}
\end{align}
%
They have the following  $q$-expansions:
\begin{align}
{\bf Y^{(2)}_3}
=\begin{pmatrix}Y_1\\Y_2\\Y_3\end{pmatrix}=
\begin{pmatrix}
1+12q+36q^2+12q^3+\dots \\
-6q^{1/3}(1+7q+8q^2+\dots) \\
-18q^{2/3}(1+2q+5q^2+\dots)\end{pmatrix}\,,
\label{Y(2)}
\end{align}
%
 where
 \begin{align}
q=\exp{ (2\pi i\,\tau)}\,.
 \label{q}
 \end{align}
%

The five modular forms of weight 4 are given as:
\begin{align}
&\begin{aligned}
{\bf Y^{(\rm 4)}_1}=Y_1^2+2 Y_2 Y_3 \ , \quad
{\bf Y^{(\rm 4)}_{1'}}=Y_3^2+2 Y_1 Y_2 \ , \quad
{\bf Y^{(\rm 4)}_{1''}}=Y_2^2+2 Y_1 Y_3=0 \ , \quad
\end{aligned}\nonumber \\
\nonumber \\
&\begin{aligned} {\bf Y^{(\rm 4)}_{3}}=
\begin{pmatrix}
Y_1^{(4)}  \\
Y_2^{(4)} \\
Y_3^{(4)}
\end{pmatrix}
=
\begin{pmatrix}
Y_1^2-Y_2 Y_3  \\
Y_3^2 -Y_1 Y_2 \\
Y_2^2-Y_1 Y_3
\end{pmatrix}\ , 
\end{aligned}
\label{weight4}
\end{align}
where ${\bf Y^{(\rm 4)}_{1''}}$ vanishes due to the constraint of Eq.\,(\ref{condition}).
%

There are  seven modular forms of weight 6:
\begin{align}
&\begin{aligned}
{\bf Y^{(\rm 6)}_1}=Y_1^3+ Y_2^3+Y_3^3 -3Y_1 Y_2 Y_3  \ , 
\end{aligned} \nonumber \\
\nonumber \\
&\begin{aligned} {\bf Y^{(\rm 6)}_3}\equiv 
\begin{pmatrix}
Y_1^{(6)}  \\
Y_2^{(6)} \\
Y_3^{(6)}
\end{pmatrix}
=(Y_1^2+2 Y_2 Y_3)
\begin{pmatrix}
Y_1  \\
Y_2 \\
Y_3
\end{pmatrix}\ , \qquad
\end{aligned}
\begin{aligned} {\bf Y^{(\rm 6)}_{3'}}\equiv
\begin{pmatrix}
Y_1^{'(6)}  \\
Y_2^{'(6)} \\
Y_3^{'(6)}
\end{pmatrix}
=(Y_3^2+2 Y_1 Y_2 )
\begin{pmatrix}
Y_3  \\
Y_1 \\
Y_2
\end{pmatrix}\ . 
\end{aligned}
\label{weight6}
\end{align}
%

The weigh 8 modular forms are nine:
\begin{align}
&\begin{aligned}
{\bf Y^{(\rm 8)}_1}=(Y_1^2+2Y_2 Y_3)^2 \, , \quad
{\bf Y^{(\rm 8)}_{1'}}=(Y_1^2+2Y_2 Y_3)(Y_3^2+2Y_1 Y_2)\, , \quad 
{\bf Y^{(\rm 8)}_{1"}}=(Y_3^2+2Y_1 Y_2)^2\, , \quad 
\end{aligned}  \nonumber\\
\\
&\begin{aligned} {\bf Y^{(\rm 8)}_3}\equiv \hskip -1 mm
\begin{pmatrix}
Y_1^{(8)}  \\
Y_2^{(8)} \\
Y_3^{(8)}
\end{pmatrix} \hskip -1 mm
=(Y_1^2+2 Y_2 Y_3) \hskip -1 mm
\begin{pmatrix}
Y_1^2-Y_2 Y_3  \\
Y_3^2 -Y_1 Y_2 \\
Y_2^2-Y_1 Y_3
\end{pmatrix} , \ \ \
\end{aligned}
\begin{aligned} {\bf Y^{(\rm 8)}_{3'}}\equiv \hskip -1 mm
\begin{pmatrix}
Y_1^{'(8)}  \\
Y_2^{'(8)} \\
Y_3^{'(8)}
\end{pmatrix} \hskip -1 mm
=(Y_3^2+2 Y_1 Y_2 ) \hskip -1 mm
\begin{pmatrix}
Y_2^2-Y_1 Y_3 \\
Y_1^2-Y_2 Y_3  \\
Y_3^2 -Y_1 Y_2 
\end{pmatrix}.
\end{aligned} \nonumber
\label{weight8}
\end{align}
%

At the fixed point  $\tau=\omega$ the modular forms take 
simple forms:
\begin{align}
&\begin{aligned}
{\bf Y^{(\rm 2)}_3}
=Y_0
\begin{pmatrix}1\\ \omega\\ -\frac12 \omega^2\end{pmatrix}\,,
\end{aligned} \qquad 
\begin{aligned} {\bf Y^{(\rm 4)}_{3}}=
\frac32 Y_0^2 
\begin{pmatrix}
1  \\
-\frac12 \omega \\
\omega^2
\end{pmatrix}\ , \qquad {\bf Y^{(\rm 4)}_1}=0 \ , \qquad
{\bf Y^{(\rm 4)}_{1'}}=\frac94 Y_0^2 \, \omega \ , 
\end{aligned} \nonumber\\
&
\begin{aligned}
{\bf Y^{(\rm 6)}_{3}}=0
\end{aligned}\ , \qquad
\begin{aligned} {\bf Y^{(\rm 6)}_{3'}}=
\frac98Y_0^3 
\begin{pmatrix}
-1  \\
2\omega \\
2\omega^2
\end{pmatrix}\ , \qquad{\bf Y^{(\rm 6)}_1}=\frac{27}{8} Y_0^3 \, , 
\end{aligned}\\
&\begin{aligned}
{\bf Y^{(\rm 8)}_{3}}=0
\end{aligned}\ , \qquad
\begin{aligned} {\bf Y^{(\rm 8)}_{3'}}=
\frac{27}{8}Y_0^4 
\begin{pmatrix}
1  \\
\omega \\
-\frac12\omega^2
\end{pmatrix}\ , \qquad {\bf Y^{(\rm 8)}_1}=0, \quad {\bf Y^{(\rm 8)}_{1'}}=0 \, , 
\quad {\bf Y^{(\rm 8)}_{1''}}=\frac94\,\omega Y_0^4 \, .
\end{aligned} \nonumber\\
& \nonumber
\label{tauomega}
\end{align}

%
\section{Modular forms at close to $\tau=\omega$}
\label{Nearby}
%
%

In what follows we present the behavior of modular forms in the vicinity of 
$\tau=\omega$. We perform Taylor expansion of modular forms  
$Y_1(\tau)$, $Y_2(\tau)$ and $Y_3(\tau)$ around $\tau=\omega$. 
We parametrize $\tau$ as:
\begin{align}
\begin{aligned}
\tau= \omega+\epsilon \, ,  
\end{aligned}
\end{align}
where  $|\epsilon|\ll 1$.
Then, the modular forms are expanded in terms of $\epsilon$.
%
{
In order to get up to 2nd order expansions of  $\epsilon$,
we parametrize the modular forms as:
\begin{align}
\begin{aligned}
\frac{Y_2(\tau)}{Y_1(\tau)}\simeq \omega\,(1+\,\epsilon_1 + k_2\epsilon^2_1 ) \, , \quad 
\frac{Y_3(\tau)}{Y_1(\tau)}\simeq -\frac{1}{2}\omega^2 \, 
(1+\, \epsilon_2 + k_3 \epsilon^2_2)\,.
\end{aligned}
\label{epST12}
\end{align}
%
These parameters are determined numerically in Taylor expansions as:}
\begin{align}
\epsilon_2=2\epsilon_1 \simeq 4.47\, i\, \epsilon\,,\qquad k_2=0.592\,,
\qquad k_3=0.546\,.
\label{ep1ep2}
\end{align}
%
The values  are obtained by using the first and second derivarives. 
The constraint $Y^2_2 + 2Y_1Y_3 = 0$ in Eq.\,\eqref{condition}
gives $\epsilon_2 = 2\epsilon_1$ and $4k_3 = 1 + 2k_2$,
which are satisfied also numerically.

%
We expres also the higher weight triplet modular forms $Y_i^{(k)}$, $k=4,6$, 
in terms of $\epsilon_1$, $k_2$ and $k_3$ using
$\epsilon_2 = 2\epsilon_1$. 
For the weight $4$ modular form we get:
\begin{align}
&  
\frac{Y_1^{(4)}(\tau)}{Y_1^2(\tau)}
\simeq \frac{3}{2}[1+\epsilon_1+ 
\frac{2}{3}\epsilon^2_1(1+\frac{1}{2}k_2 + 2k_3)] \, ,
\\ 
&\frac{Y_2^{(4)}(\tau)}{Y_1^2(\tau)} \simeq \omega\,
\left( -\,\frac{3}{4} + \epsilon^2_1(1+2k_3-k_2)\right)\, ,
\\
&\frac{Y_3^{(4)}(\tau)}{Y_1^2(\tau)} \simeq \omega^2\,  
\left( \frac{3}{2} + 3\epsilon_1+ \epsilon^2_1(1+2k_2+2k_3)\right )\,.
\label{epST43}
\end{align}
%

In a similar way we get the expressions for the two relevant 
weight $6$ modular forms:
\begin{align}
&  \begin{aligned}
\frac{Y_1^{(6)}(\tau)}{Y_1^3(\tau)} \simeq -\,3\epsilon_1 
-\,\epsilon^2_1(2+k_2+4k_3)\,,
\end{aligned} 
\\
&  \begin{aligned}
\frac{Y_2^{(6)}(\tau)}{Y_1^3(\tau)} \simeq
-\,\omega\, [3\epsilon_1+\epsilon^2_1(5+k_2+4k_3)]\,  ,
\end{aligned}
\\
&  \begin{aligned}
\frac{Y_3^{(6)}(\tau)}{Y_1^3(\tau)} \simeq
\omega^2\, [\frac{3}{2}\epsilon_1+\frac{1}{2}\epsilon^2_1(8+k_2+4k_3)] ,
\end{aligned}
\\
&  \begin{aligned}
\frac{Y_1^{'(6)}(\tau)}{Y_1^3(\tau)} \simeq 
-\,\left (\frac{9}{8} + \frac{15}{4}\epsilon_1 + 
\epsilon^2_1(\frac{7}{2} + k_2 +\frac{11}{2}k_3)\right )\,, 
\end{aligned} 
\\
&\begin{aligned}
\frac{Y_2^{'(6)}(\tau)}{Y_1^3(\tau)}  \simeq
\omega\,
\left (\frac{9}{4}+3\epsilon_1+\epsilon^2_1(1+2k_2+2k_3)\right )\,,
\end{aligned}
\\
&  \begin{aligned}
\frac{Y_3^{'(6)}(\tau)}{Y_1^3(\tau)}  \simeq
\omega^2\,
\left (\frac{9}{4}+\frac{21}{4}\epsilon_1+
\epsilon^2_1(4+\frac{17}{4}k_2+2k_3)\right )\, .
\end{aligned}
\label{epST6p3}
\end{align}

\section{A measure of fit}
\label{fit}

 As a measure of goodness of fit, we use the sum of one-dimensional 
$\Delta\chi^2$ for eight   observable quantities
$q_j=(m_d/m_b, \, m_s/m_b,\,m_u/m_t, \, m_c/m_t,\, |V_{us}|,\, |V_{cb}|,\, |V_{ub}|,\, \delta_{CP})$.
By employing the Gaussian approximation, we difine
$N\sigma\equiv \sqrt{\Delta \chi^2}$, where
\begin{align}
\Delta \chi^2=
\sum_j \left ( \frac{q_j-q_{j,{\rm best\,fit}}}{\sigma_j}\right )^2\,.
\end{align}
%

\vskip 1 cm


\end{document}